\documentclass{aa}  
\usepackage{graphicx}
\usepackage{natbib}
\usepackage{float}
\usepackage{xcolor}
\usepackage{txfonts}
\begin{document} 
\title{Massive and old quiescent galaxies at high redshift}

   \author{Giacomo Girelli
          \inst{1,2}
          \thanks{giacomo.girelli2@unibo.it}
          \and Micol Bolzonella\inst{1}
          \and Andrea Cimatti\inst{2,3}
          }
    \authorrunning{Girelli et al. }
   \institute{
     INAF – Osservatorio di Astrofisica e Scienza dello Spazio di Bologna, via P. Gobetti 93/3, I-40129 Bologna, Italy
     \and
     Universit\`a di Bologna, Dipartimento di Fisica e Astronomia, Via P. Gobetti 93/2, I-40129, Bologna, Italy
     \and
     INAF - Osservatorio Astrofisico di Arcetri, Largo Enrico Fermi 5, I-50125 Firenze, Italy 
   }

   \date{Received XXX; accepted YYY}
 
  \abstract
   {}
{Questions of how massive quiescent galaxies rapidly assembled and how abundant they are at high redshift are increasingly important in the study of galaxy formation. 
Looking at these systems can shed light on the processes of galaxy mass 
assembly and quenching of the star formation at early epochs. In order 
to address these questions, we aim to identify and characterize massive quiescent galaxies from $z \sim 2.5$ out to the highest redshifts at which these systems can be found. The final purpose is to compare the results with the predictions of state-of-the-art semi-analytical models of galaxy formation and evolution.}
{We defined observer-frame color-color diagrams to optimally select quiescent galaxies at $z>2.5$ and applied them to the COSMOS2015 catalog. We refined the spectral energy distribution (SED) fitting analysis for the selected candidates to confirm their quiescent nature, then derived their number density, mass density, and stellar mass functions. Finally, we compared the results with previous observations and some current semi-analytic models.}
{We selected candidates for quiescent galaxies in the redshift range $2.5\lesssim z\lesssim 4.5$ from the COSMOS2015 catalog by means of two color-color diagrams.  The additional SED fitting analysis allowed us to select  $128$ galaxies, consistent with being massive ($\log(M_*/M_\odot)\geq 10.6$),  old (ages $\gtrsim 0.5\,{\rm Gyr}$), and quiescent ($\log({\rm sSFR \, [{yr}^{-1}}])\leq -10.5$) objects at high redshift ($2.5<z<4.5$). Their number and mass densities are in fair agreement with previous observations and, if confirmed, show a discrepancy with current semi-analytical models of galaxy formation and evolution \citep{Henriques15}, that underpredict the number of massive quiescent systems up to a factor of $\sim 12$ at $2.5\leq z<3.0$ and $\sim 10$ at $z\sim 4.0$. The evolution of the stellar mass functions (SMFs) of these systems is similar to previous estimates and indicates a disagreement with models, particularly with regard to the shape of the SMF.}
{The present results add further evidence to the possibility that  
massive and quiescent galaxies can exist out to at least $z \sim 4$. If future spectroscopic observations carried out with, for example, the James Webb Space Telecope (JWST), confirm the substantial presence of such a population, further work on modeling the stellar mass assembly, as well as supermassive black hole (BH) accretion and feedback processes at early cosmic epochs, is needed to understand how these systems formed, evolved, and quenched their star formation.}


   \maketitle


\section{Introduction}
\label{intro}

The formation of galaxies and their properties are driven by the evolution of both dark matter and baryons. While dark matter halos assembled most of their mass through sequential merging, some fundamental questions remain still open \citep{Naab17}, particularly those regarding the formation of massive galaxies and their transformation into quiescent systems.

Recent developments support a scenario where galaxies assembled most of their mass not only through sequential merging, but also through smooth accretion by cold gas streams that penetrate the shock-heated media of massive dark matter halos and grow dense, unstable, turbulent discs with bulges, and trigger rapid star formation (e.g., \citealt{Dekel09}).

Different models of galaxy formation and evolution have been developed in order to explain the emergence of massive objects (e.g., \citealt{Sheth01, Henriques15}) and their transformation into quiescent systems (e.g., \citealt{Guo11, Guo13, Naab14,Henriques17}). However, when these models are applied to simulations (e.g., the Munich Simulation: \citealt{Kauffmann99,Springel01,Croton06,Delucia07,Henriques15,Croton16}), they predict results which are not fully consistent with the observed evolution of galaxy properties. In particular, the presence of a population of high-redshift massive (and sometimes quiescent) galaxies can create tension between models and observations (e.g., the impossibly early galaxy problem, as in \citealt{Steinhardt16}, referring to the problem of such rapid assembly of massive systems). Therefore, a robust determination of the abundance of massive quiescent galaxies at high redshift is a powerful test bench for galaxy formation models that also have to reproduce the mechanisms for quenching  star formation in order to produce quiescent galaxies.

Many recent studies (e.g., \citealt{Daddi04,Ilbert06,Fontana06,Wiklind08,Caputi12,Castro12,Stefanon13,Muzzin13,Nayyeri14,Straatman14,Mawatari16,Wang16,Merlin18,Deshmukh2018,Merlin2019}) have identified a population of massive quiescent galaxies at high redshift ($z \gtrsim 2.5$). Due to the current observational limits, these objects are just photometric candidates based on spectral energy distribution (SED) fitting analysis of broad-band photometry with population synthesis models. There are few notable exceptions for which deep near-infrared (NIR) spectroscopy confirms the passive nature of the candidates and determines their redshifts (at $z = 3$ in \citealt{Gobat12}, and $z = 3.7$ in \citealt{Glazebrook17}). Moreover, in a recent work \citet{Schreiber18} obtain Keck–MOSFIRE $H$ and $K$-band spectra for $24$ candidate quiescent galaxies at $3 < z < 4$ and confirm the redshift and passive nature for eight of them, adding further evidence to the existence of a significant population of quiescent galaxies at high-redshift. This is a strong evidence that star formation in these galaxies occurred very fast within the first billion years of the universe, followed by a passive evolution \citep{Mancini09,Stefanon13,Straatman14,Oesch13}. 
The common technique adopted to select and identify quiescent galaxies at $z > 3$ consists of two steps: the selection of likely candidates based on colors (observed or rest-frame) and the subsequent identification of the most likely old and high redshift galaxies among these candidates by means of a spectral energy distribution (SED) fitting analysis with population synthesis models (e.g., \citealt{Wiklind08, Nayyeri14,Wang16}). This two-step process is required because, as it is discussed further in this paper, the colors of these high redshift quiescent galaxies are to some extent degenerate with regard to those of dusty star-forming galaxies at the same or lower redshift (as is also known for quiescent galaxies at lower redshift, e.g., \citealt{Pozzetti00}). A recent example is the work by \citet{Merlin18} in which the authors search for passive galaxies at $z>3$ in the GOODS-South field using photometric data and then select galaxies with star formation rate ${\rm SFR}=0$ with the aid of a SED fitting analysis adopting top-hat star formation histories. All these results demonstrate the increasing importance of the quest for identifying the population of passive and quiescent galaxies at high redshift.

The aim of this paper is to devise observed color-selection diagrams that can efficiently identify the bulk of the population of $z \ge 2.5$ massive quiescent galaxies, including those that could be missed by other selection techniques. 
In Sect.~\ref{colselection}, we present new observed color selections to identify quiescent galaxies at $z\gtrsim 2.5$.  We carefully studied the effects of all the parameters characterizing evolutionary tracks, which is presented in Appendix~\ref{appendix1}. The data used to test the new criteria are given in Sect.~\ref{cosmos}. 
In Sect.~\ref{candidates}, we present the selection of the quiescent candidates through color selections and in Sect.~\ref{SEDfitting}, the SED fitting analysis needed to ascertain their passive nature. In Sect.~\ref{numdens}, we revise the number density of quiescent objects at high redshift and compare our results to recent galaxy formation models. Moreover, in Sect.~\ref{massfct}, we present our estimate of the mass functions for the quiescent population. A discussion on our results and a comparison with previous observations and current semi-analytic models is presented in Sect. \ref{discussion}. A summary of the methods adopted and the results is presented in Sect. \ref{concl}. 
Throughout this paper, a standard cold dark matter ($\mathrm{\Lambda CDM}$) cosmology is adopted with Hubble constant $H_0=70\, \mathrm{km}\, \mathrm{s}^{-1}\,\mathrm{Mpc}^{-1}$, total matter density $\Omega_{\mathrm{M}} = 0.3$ and dark energy density $\Omega_\Lambda = 0.7$. All magnitudes are expressed in the AB system and log is base $10$ logarithm, if not otherwise specified.


\section{Color selections at $z>2.5$}
\label{colselection}

One of the most efficient ways of selecting quiescent galaxies is based on colors, a method which relies on the identification of spectral features via photometry. 

Color-color selection can be performed using rest-frame or observed quantities. The approach with rest-frame colors, such as $UVJ$ \citep{Williams09} and ${\rm NUV}RJ$ (see \citealt{Laigle16} or \citealt{Ilbert13} for details), requires an estimate of the redshift, which for large and deep surveys is usually derived from photometry, with rest-frame colors evaluated using the observed SED. High-redshift quiescent galaxies are characterized by SEDs which make them often too faint to be detected at optical wavelengths, and therefore rest-frame colors are usually extrapolated \citep{Ilbert13}. Instead, apparent colors (e.g., \citealt{Nayyeri14}) have the advantage of relying only on observations, although the lack of redshift measurement can add a further level of degeneracy. In this paper we will follow this second approach, in order to pre-select quiescent candidates based only upon observed data, and then an ad-hoc SED fitting analysis is performed in order to confirm their nature.

To identify quiescent galaxies by means of observed colors we use the strength of the Balmer and D$4000$ breaks (e.g.,~\citealt{Bica94}), redshifted to near-infrared (NIR) wavelengths for $2.5 \lesssim z \lesssim 4.5$. The Balmer break at $\lambda_{\rm rest}=3646$\,\AA\ is due to the absorption by Balmer series down to the Balmer limit (strongest in A-type stars). The so-called D$4000$ break at $4000$\,\AA\, is mostly produced by H and K absorption lines of calcium at $\lambda_{\rm rest}=3968$\,\AA\ and $\lambda_{\rm rest}=3933$\,\AA\ respectively, characteristic of solar-type stars, along with several metallic absorption lines. Both discontinuities can be used to select galaxies which ceased star formation (post-starburst, quiescent passive) since their breaks are much more prominent than in star forming galaxies.

\subsection{Color predictions with evolutionary tracks}

To identify the locus populated by high-$z$ passive galaxies on a color-color diagram we used evolutionary tracks derived with stellar population synthesis (SPS) models. Some previous work on color selection determination (e.g., \citealt{Nayyeri14,Wiklind08, Wang16, Mawatari16}) adopts a fixed age for the stellar populations and follows their color evolution with redshift. In our work, we decided to follow a more physically-motivated approach, letting colors evolve also with age after having fixed a formation redshift in order to avoid ages larger than the age of the universe at the considered redshift. Therefore, in our approach we have to consider several values of formation redshift to explore the whole parameter space of the evolutionary tracks, as done in Appendix~\ref{sect_zf}.

To parametrize quiescent objects, we computed color tracks from BC03 \citep{B&C03} models, adopting a Chabrier IMF \citep{Chabrier03}, solar metallicity ($Z=Z_{\odot}=0.02$), and exponentially declining star formation histories with different $e$-folding times $\tau$. 
Star-forming galaxies are generally assumed to be reproduced by a constant star formation template: we built evolutionary tracks of this galaxy template using the code \emph{fsps} \citep{Conroy09,Conroy10}, including the contribution of nebular emission lines that are evaluated with the results of \citet{Byler17}  based on {\sc CLOUDY} \citep{Ferland13}. 
Among the free parameters available in the model, we chose to adopt the Chabrier IMF, Padova's stellar isochrones, and BaSeL libraries for better consistency with BC03 models. 
We considered the aging of both the star-forming and the quiescent templates. Therefore, a redshift of formation must be fixed when computing the color tracks as a function of redshift, and we chose as our reference $z_{\rm form}=6$.
We also allowed for different values of dust extinction $A_V$  applied through the Calzetti's attenuation law \citep{Calzetti00}. 

We tested the robustness of the selection criteria against different choices of parameters for extinction laws, initial mass functions (IMFs), star formation histories (SFHs), metallicities, stellar population synthesis (SPS) models (e.g., \citealt{Maraston05}, hereafter M05), redshift of formation, and also considering whether the inclusion (or not) of emission lines changes the selection criteria in Appendix~\ref{appendix1}.

We systematically looked for optimal selections using near- and mid-infrared observed colors. It is indeed necessary to use red bands and/or near/mid-infrared bands for three main reasons: first, because quiescent galaxies are characterized by intrinsically red SEDs; second, because the effect of redshift and $k$-correction makes these objects almost undetected in optical blue bands due to their faint UV emission; and finally, because at $z>3$ , the radiation blue-ward the Lyman-limit is largely suppressed by the inter-galactic medium (IGM) absorption and, therefore, all the galaxies start to become suppressed in the $u$ filter \citep{Guhathakurta90,Steidel96,Steidel99}. In particular, we used the photometry in the bands $i$, $z$, $Y$, $J$, $H$, $K_{\rm s}$, IRAC$[3.6]$ and IRAC$[4.5]$, with transmission curves of the instruments which have observed the COSMOS field\footnote{ Transmission curves, including filter, optics, mirror, atmosphere, and detector are taken from  http://cosmos.astro.caltech.edu/page/filterset}, and whose characteristics are listed in Table~\ref{table:1}. 

Based on the vast exploration of possible colors, we selected two criteria that perform well in selecting quiescent objects at $2.5 \lesssim z\lesssim 4.5$.
We stress that the purpose of this work is to be as inclusive as possible: we decided to maximize the completeness of the quiescent galaxies selection by including objects that can be missed by other criteria. The consequence of this choice will be the lower purity of the sample; contamination from other types of objects will have to be removed with the aid of an SED fitting.

\begin{table}
\caption{List of filters used in this work and their main characteristics. The last two columns represent the median limiting magnitude with a $3''$ diameter aperture and a depth of $2\sigma$, i.e., $S/N=2$ and $3\sigma$, i.e., $S/N=3$, respectively.}
\label{table:1}     
\centering
\begin{tabular}{c c c c l l }      
\hline\hline
Filter & $\lambda_{\rm eff}$ (\AA) & $\Delta\lambda$ (\AA) & $2\sigma (3'')$ & $3\sigma (3'')$\\ \hline
$u_{\rm CFHT}$       & $3823.3$  & $670$   & 27.8 & 26.6\\
$B_{\rm Subaru}$     & $4458.3$  & $946$   & 27.8 & 27.0\\
$V_{\rm Subaru}$     & $5477.8$  & $955$   & 26.9 & 26.2\\
$r_{\rm Subaru}$     & $6288.7$  & $1382$  & 26.7 & 26.5\\
$i^{+}_{\rm Subaru}$ & $7683.9$  & $1497$  & 26.5 & 26.2\\
$z^{++}_{\rm Subaru}$& $9105.7$  & $1370$  & 26.0 & 25.9\\
$Y_{\rm VISTA}$      & $10214.2$ & $970$   & 25.5 & 25.3\\
$J_{\rm VISTA}$      & $12534.6$ & $1720$  & 25.2 & 24.9\\
$H_{\rm VISTA}$      & $16453.4$ & $2900$  & 24.8 & 24.6\\
$K_{{\rm s},{\rm VISTA}}$  & $21539.9$ & $3090$  & 24.8 & 24.7\\
IRAC$[3.6]$          & $35634.3$ & $7460$  & 26.3 & 25.5\\
IRAC$[4.5]$          & $45110.1$ & $10110$ & 26.0 & 25.5\\
\hline                  
\end{tabular}
\end{table}

\subsection{$J K_s [3.6] [4.5]$ selection}
\label{JK3645}
 
We designed the color selection using the observed colors $J-K_{\rm s}$ and ${\rm IRAC}[3.6] - {\rm IRAC}[4.5]$ (hereafter $[3.6]-[4.5]$) to select quiescent galaxies in the redshift range $z\approx 2-4$.  At $z\approx 2$ the Balmer break is redshifted at $\lambda\approx 1.09\,\mu$m, while for $z\approx 4$ to $\lambda\approx 1.82\,\mu$m; the D$4000$ break is redshifted at $\lambda\approx 1.2\,\mu$m at $z\approx 2$ and at $\lambda\approx 2.0\,\mu$m at $z\approx 4$. The color $J-K_{\rm s}$ can therefore identify the breaks, while the color $[3.6]-[4.5]$ has been chosen since it best complements the other color to avoid overlaps between different tracks.

In the top panels of Fig.~\ref{1}, two SEDs of quiescent galaxies at fixed ages and redshifted to different $z$ with no dust extinction are shown to illustrate the location of the D$4000$ and Balmer breaks for the objects of interest with respect to the used filters. The cyan line shows a SED of a quiescent galaxy of $0.5$ Gyr redshifted to $z=4$ while the magenta line shows a SED of a quiescent galaxy of $1$ Gyr at $z=3$. Filter transmission curves are also shown in order to indicate the location of the Balmer and D$4000$ breaks with respect to the chosen colors. 

We identified the following criteria to select quiescent galaxies in the redshift range $z=2-4$ and limit the contamination from star-forming galaxies:
\begin{equation}
\begin{cases} 
([3.6]-[4.5])\ge 0 \\
(J-K_{\rm s})\ge 0.975([3.6]-[4.5])+1.92 \mbox{ for } [3.6]-[4.5]<0.4\\ 
(J-K_{\rm s})\ge 9.0([3.6]-[4.5])-1.33 \mbox{ for } [3.6]-[4.5]\ge 0.4
\end{cases}
\end{equation}
which correspond to the gray shaded area in the left panel of Fig.~\ref{1}. In the same panel, we show the color-color diagram and different evolutionary tracks built as described in the previous section. 
More precisely, we show three different tracks parametrizing quiescent galaxies with different extinction values. In particular, magenta tracks show a track with $A_V=0$, orange a track with $A_V=1$ and the red a track with $A_V=2$. The choice of $A_V=2$ is an extreme case: in fact, we do not expect to find any quiescent candidate with such high extinction value. However, in order to be as inclusive as possible, we do not exclude the region covered by this track. 
High-extinction star-forming galaxies can, in principle, contaminate the sample, but even though the black track in Fig.~\ref{1}, which reproduces the color evolution of a galaxy with constant star formation history and $A_V=5$, falls within the selection region, these interlopers are at lower redshift ($z\lesssim 2$) and can be removed from the sample once SED fitting is performed.

\subsection{$H K_s [3.6]$ selection}
\label{HK36}

\begin{figure*}[ht]
   \centering
   \includegraphics[width=18cm]{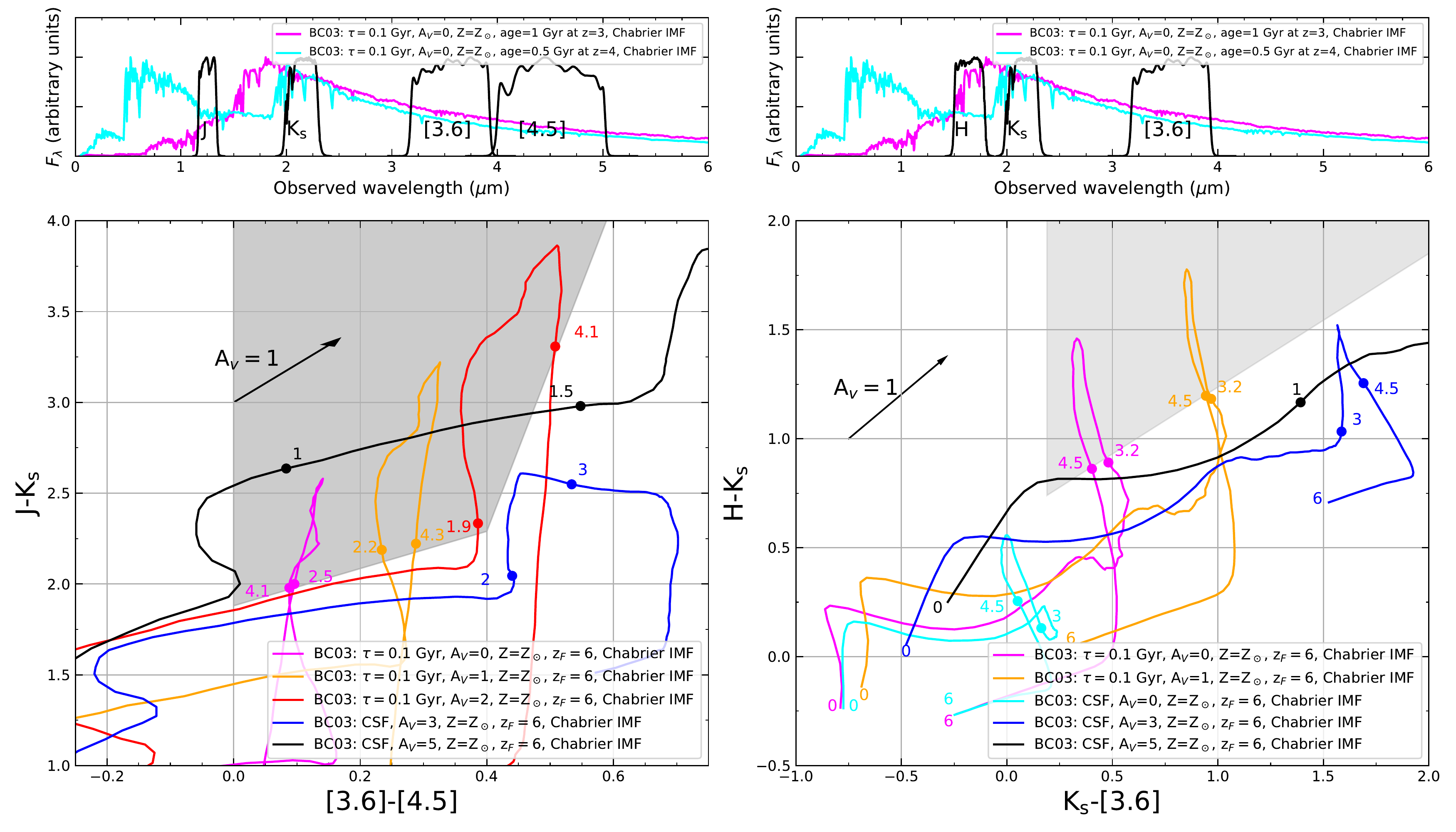}
   \caption{\textbf{Top panels:} two SEDs of quiescent galaxies. SED in magenta represents a population of $1$\,Gyr redshifted to $z=3$ (i.e., $z_{\rm form} \approx 6$), while the cyan line shows the SED of a population of $0.5$\,Gyr redshifted to $z=4$ (i.e., again $z_{\rm form}\approx 6$). Filters transmission curves used in COSMOS field and in the tracks evaluations are also shown. \textbf{Bottom panels:} the $JK_{\rm s} [3.6][4.5]$ and $HK_{\rm s}[3.6]$ color-color diagrams with different evolutionary tracks both for star-forming galaxies and passive galaxies. Tracks representing star-forming galaxies are shown in cyan, blue and black with $A_V=0$, $A_V=3$, and $A_V=5$ respectively. Three tracks for quiescent galaxies (with exponentially declining star formation history $\tau=0.1$\,Gyr and solar metallicity) are shown in magenta, orange and red with different extinction values ($A_V=0$, $A_V=1$, and $A_V=2$ respectively). Gray shaded areas represent the selection region for quiescent galaxies at $2.5 \lesssim z \lesssim 4$ in the left panel and at $3 \lesssim z \lesssim 4.5$ in the right panel. The numbers near the tracks of the same color represent the redshift. A vector corresponding to a magnitude extinction of $A_V=1$ using Calzetti's law is also shown.}
              \label{1}%
    \end{figure*}

This color selection is designed to identify quiescent galaxies at slightly higher redshift with respect to the previous color selection, that is, in the range $z=3-4.5$, using filters $H$, $K_{\rm s}$ and IRAC $3.6\,\mu$m.
From $z\approx 3$ to $z\approx 4.5,$ both the Balmer and the D$4000$ break are located in the wavelength range bracketed by $H$ and $K_{\rm s}$ filters (from $\lambda=1.64 \,\mu$m to $\lambda=2.15\,\mu$m). Since the D$4000$ and Balmer breaks develop on different timescales, the color  $H-K_{\rm s}$ can be used to select quiescent galaxies for broad ranges of age. In the top-right panel of Fig.~\ref{1}, two SEDs of quiescent galaxies at fixed ages and redshifted to different $z$ are shown to illustrate the location of the breaks with respect to the filters used.

In the bottom-right panel of Fig.~\ref{1}, we show the color-color diagram and different evolutionary tracks built as described in the previous section.  As for the previous color selection, the color $K_{\rm s}-[3.6]$ has been chosen in order to best complement the information of the other color.

To select quiescent galaxies in the redshift range $z=3-4.5$, and to exclude most star-forming galaxies, we designed the following selection criteria:
\begin{equation}
\begin{cases} 
K_{\rm s}-[3.6]\ge 0.2\\ 
H-K_{\rm s}\ge 0.617 (K_{\rm s}-[3.6])+0.58 
\end{cases}
\end{equation}
which correspond to the gray shaded area in the right panel of Fig.~\ref{1}. Star-forming galaxies may enter the color selection region for extreme values of extinction ($A_{V}\gtrsim 5$) at lower redshift: the contamination by dusty and star-forming galaxies cannot be completely excluded, but they can be removed with the SED fitting analysis.

\section{Application to real data}
\label{cosmos}

We applied the proposed color-color diagrams to real data. To this aim, we used the multi-waveband deep photometric observations available in the COSMOS field \citep{Scoville07}.

\subsection{The COSMOS2015 catalog and colors}

We used the COSMOS2015 catalog \citep{Laigle16}, which is the latest public release of data in the COSMOS field. This catalog provides the deepest optical and infrared observations of the field and the photometry has been obtained using a detection image evaluated by combining NIR images of UltraVISTA ($YJHK_{\rm s}$) with the optical $z^{++}$-band data from Subaru. Since deep optical and IR data are especially important for the purpose of this paper, the characteristics of the catalog are particularly suitable for our work. In COSMOS2015, optical and NIR photometric data are provided in different apertures: $2''$, $3''$ (denoted APER2 and APER3 respectively), ISO (isophotal), and AUTO apertures. AUTO apertures are object to object variable apertures, automatically evaluated by SExtractor \citep{Bertin96} in order to approximately measure the total flux of the object. IRAC data are instead provided only as total magnitudes and fluxes.

In the color-color diagrams presented in Sects.~\ref{HK36} and \ref{JK3645}, we combine NIR and IRAC data and, therefore, total magnitudes are needed for all the filters. Following \citealt{Laigle16}, instead of using AUTO magnitudes in NIR filters, we rescaled APER3 magnitudes to reproduce the total flux normalization for each object since these aperture magnitudes provide better results in SED fitting for photometric redshifts. We computed total magnitudes in optical and NIR bands following the prescriptions in Eqs.~9 and 10 of \citealt{Laigle16}. In particular, we applied to each object a photometric offset that normalizes the aperture magnitudes to the total ones, preserving the colors that have been better determined in aperture photometry. We also corrected magnitudes for foreground galactic extinction using the values given in the catalog, which are computed at each object position using the \citet{Schlegel98} dust maps. Finally, we applied corrections to both NIR and IRAC magnitudes using the small systematic offsets given in the catalog, estimated by matching the predicted magnitudes and the observed ones using the spectroscopic sample.

A census of the photometric bands adopted in this paper and the depth of observations at $2\sigma$ and $3\sigma$ are presented in Table ~\ref{table:1}. The $3\sigma$ values are provided in \citet{Laigle16}, while $2\sigma$, that will be used in the following, have been evaluated as the magnitude of objects with photometric error $\Delta m\sim 0.54$ (corresponding to $S/N=2$).

\subsection{Parent sample}
\label{parent}

\begin{figure}[ht]
\centering
\includegraphics[width=9cm]{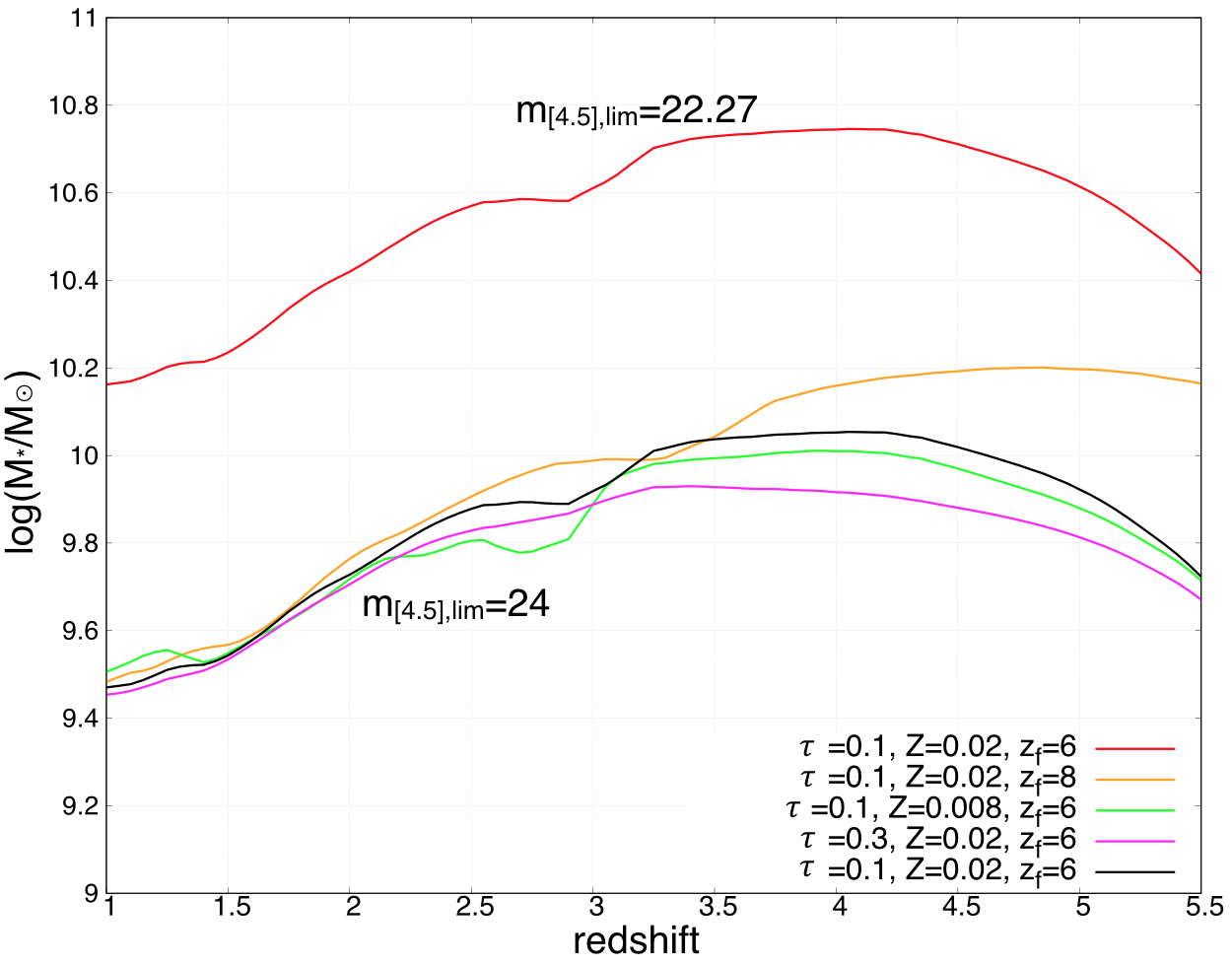}
\caption{Stellar mass as a function of redshift for different theoretical parametrizations of quiescent galaxies: we adopted an exponentially declining star formation history with characteristic time $\tau =0.1$\,Gyr, $Z=Z_{\odot}=0.02$, $A_V =0$, $z_{\rm form} =6$. Red and black lines correspond to magnitudes $m_{[4.5]}=22.27$ and $24$, respectively. Orange, green, and magenta lines represent different models of a quiescent galaxy differing in redshift of formation, metallicity, and $e$-folding time of the SFH, respectively.}
\label{mlim}%
\end{figure}

\begin{figure*}[ht]
\centering
\includegraphics[width=18cm]{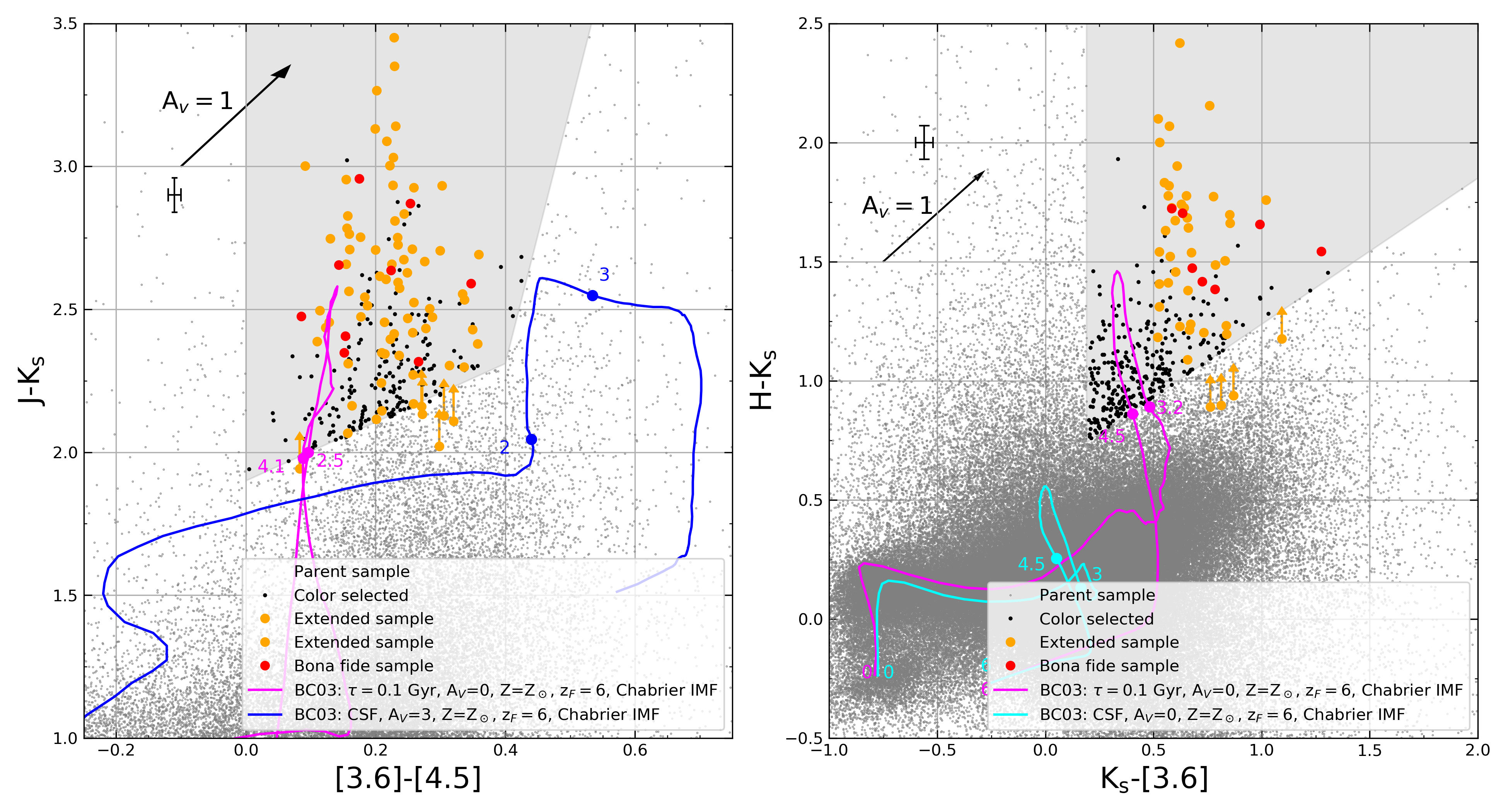}
\caption{Observed color-color diagrams. Left: $JK_{\rm s}[3.6][4.5]$. Right: $HK_{\rm s}[3.6]$. Gray shaded areas represent the color selection regions defined in Sect.\ref{colselection} and the evolution of the colors of a quiescent and a star-forming galaxies are plotted as a reference. Gray points represent the whole parent sample selected as in Sect.~\ref{parent}, black points show the color-selected-only quiescent candidate. Red points are "bona fide" objects, while orange points belong to the "extended" sample, both defined in Sect.\ref{SEDfitting}.  The mean photometric errors of the color-selected candidates (black points) are also shown. Objects that are non-detected in one or more filters in the color-selected sample are not shown in order to improve clarity given their large number. Only undetected objects belonging to extended or  bona fide samples are shown with arrows indicating the lower limits of the involved color. Non-detections correspond to $<2\sigma$.}
\label{cand}
\end{figure*}

In this work we focus on massive galaxies and in order to build a mass-selected sample, we used as a proxy NIR apparent magnitudes in filter IRAC$[4.5]$. In fact, since stars with small masses ($M\lesssim 1$ $M_{\odot}$), which make up the bulk of the mass of a galaxy, mostly emit in the redder part of the spectrum, a selection in NIR magnitudes roughly corresponds to a selection in stellar mass. Moreover, as shown e.g., in \citet{Bell01}, the stellar mass-to-light ratios in the rest-frame NIR bands vary only by a factor of $2$ or less over a wide range of star-forming histories, in contrast with a factor of $10$ in blue bands. This means that the luminosity of a galaxy in the NIR is a good tracer of its stellar mass. 

Using population synthesis models, we derived the relations between stellar mass and magnitudes in filter IRAC$[4.5]$ as a function of redshift: at  $z=3$ the expected magnitude of a passive galaxy with mass $M= 10^{10.6}\, M_{\odot}$ is $m_{[4.5]}\simeq 22.3$, that is, passive galaxies brighter than this limit should be characterized by stellar masses larger than $10^{10.6}\, M_{\odot}$. 
The same type of object with the same stellar mass can be observed with a magnitude of $m_{[4.5]}\simeq 22.6$ at $z=4$ and $m_{[4.5]}\simeq 22.3$ at $z=5$. Assuming a different formation redshift $z_{\rm form}=8$, the expected magnitude of a passive galaxy with $M= 10^{10.6}\, M_{\odot}$ at $z=3$ is $m_{[4.5]}\simeq 22.4$, while considering models with low metallicity ($Z=0.008$) or a longer timescale of star formation ($\tau =0.3$ Gyr), the corresponding magnitude is $m_{[4.5]}\simeq 22.2$. 

Based on previous considerations, we first defined and selected a parent sample with a cut at $m_{[4.5]}\le 24$ to guarantee both the sensitivity to stellar masses $\log (M/M_\odot) \gtrsim 10$ for quiescent galaxies and reliable photometry. In terms of the signal-to-noise ratio, this magnitude limit ensures that more than $99\%$ of the considered data have an $S/N>3$ (also see  Table~\ref{table:1}). Moreover, we considered only sources with $S/N\ge 3$ (i.e., $\Delta m\le 0.36$) at $4.5\,\mu$m. Spectroscopically confirmed stars and active galactic nuclei (AGN), for a total of $743$ objects, have also been removed. 

All the above selections were performed inside the region with the best quality photometry: FLAG\_PETER, which defines good optical areas, that is, masking regions where bright stars may contaminate the photometry of nearby objects, while FLAG\_COSMOS and FLAG\_HJMCC define the area in COSMOS field covered by UltraVISTA. UltraVISTA data are primarily important in this work as they are the basis of our selections. The final sky area considered after applying the cited flags is $1.38\,\mathrm{deg}^2$. 

The parent sample is then composed of $212\,897$ objects (out of the $1\,182\,108$ entries in the original COSMOS2015 catalog). The whole parent sample is shown in gray in Fig.~\ref{cand} in our color-color diagrams.

\subsection{Photometric requirements}
\label{candidates}

To select quiescent candidates through color-color diagrams, we impose the following photometric requirements:

\begin{enumerate}
\item We defined as the detection limit in each filter of interest a conservative value of  $S/N=2$, corresponding to a maximum photometric error of $\Delta m=0.54$ magnitudes, in order to include in the color selection any possible quiescent candidate. In addition, we also considered  galaxies non-detected (i.e., with $S/N < 2$) in some of the filters used in the diagrams; when a non-detection occured, the magnitude limit in that filter was considered from Table~\ref{table:1} to derive a lower or upper limit of the color. 

\item To minimize the fraction of lower redshift and star-forming objects, we removed from our sample those objects detected in the $u$ band with $S/N > 2$; given our parent sample cut at $m_{[4.5]}\le 24$, this choice corresponds to removing from the sample those objects with blue  $u-m_{[4.5]}$ colors, which are characteristic only of star-forming galaxies or very-low-redshift quiescent galaxies. 

\item We imposed a non-detection condition at $24\,\mu$m measured by MIPS (Multiband Imaging Photometer on Spitzer) to minimize the fraction of dusty star-forming galaxies that would significantly emit at $24\,\mu$m because of the dust re-emission; a detection limit of $S/N=2$ was chosen.
\end{enumerate}

\subsection{Candidates identification through color selections}

Candidates are identified from the parent sample using color-color diagrams defined in Sect.~\ref{colselection}. As mentioned above, we also include in the selection extremely faint quiescent objects that may not be detected
in some of the filters involved in the diagram. Moreover, sources which satisfy the selection criteria but do not meet the $u$ and/or the $24\,\mu$m non-detection conditions were removed from the sample (a total of $182743$ objects in the parent sample were rejected due to these cuts).

A total of $1047$ objects ($0.49\%$ of parent sample) were selected using at least one of the two criteria ($183$ objects were selected by both color selections simultaneously. See Table~\ref{table:2} for details).
In Fig.~\ref{cand}, we show these candidates as black dots, together with two evolutionary tracks for comparison with previous plots. Objects that are non-detected in one or more filters in the color-selected sample are not shown for the purposes of clarity given their large number. We notice that a large number of objects is located just below the selection boxes: the color errors will therefore result in a larger number of objects scattered upwards into the selection box than downwards, producing an overestimate of the number of selected objects. This bias and the other possible sources of contamination are the object of the SED fitting refinement illustrated in Sect.~\ref{SEDfitting}. 

Figure~\ref{prop} illustrates the properties of the selected objects in gray: the redshift distribution of the selected sample peaks at $z_{\rm phot}\approx 3.5$ and appears to be bimodal, as a result of the combination of the two adopted color diagrams. A non-negligible fraction of objects exhibits photometric redshifts in the range $2<z_{\rm phot}<2.5$: a contamination by non-quiescent and/or $z<2.5$ objects was expected. In fact, the chosen color selections (with generous selection boxes), the parent sample (with a deep cut at 24 magnitudes at $4.5\,\mu$m), and the photometric requirements (with an undetection requirement at $2\sigma$ in $u$ and $24\,\mu$m bands) have been designed in order to include any possible quiescent candidate, given their rareness at $z\ge 2.5$.  We also expect a certain degree of contamination by star-forming galaxies with prominent emission lines. With the aid of the full photometric data, the SED fitting procedure is expected to be able to identify the truly quiescent objects at the highest redshifts and remove the interlopers belonging to different redshift ranges or galaxy types.

\section{SED fitting and physical properties}
\label{SEDfitting}

The SED fitting analysis of the selected objects allows us to select the most massive ($\log (M_*/M_\odot) >10.6$) high redshift ($z\gtrsim 2.5-3$) quiescent galaxies, hereafter defined as galaxies characterized by $\log ({\rm sSFR\, [\,yr^{-1}]}) <-10.5$, where sSFR is the specific star formation rate (${\rm sSFR}={\rm SFR}/M_*$). Since there are no spectroscopic redshifts measured for the candidates selected in the previous section, to evaluate the physical properties of the candidates we assumed the photometric redshifts that have been carefully optimized by \citet{Laigle16} in the COSMOS2015 catalog. We then carried out the SED fitting to estimate physical properties, such as the stellar mass, extinction, age, and specific star formation rate (sSFR). The exploration of the parameter space has not been designed to consider all the possible SEDs, but only to break the degeneracy between the two specific galaxy populations we expect from the color-color selection, that is, high-redshift quiescent objects and dusty star-forming contaminants.

We used the hyperzmass code\emph{} \citep{Bolzonella00,Bolzonella10} to evaluate the best fit SED, corresponding to the minimum $\chi^2$ derived from the comparison between observed and model photometry at fixed $z=z_{\rm phot}$ from COSMOS2015. From the best fit SED we derive the stellar mass, star formation rate, age, and extinction. In the following we describe the relevant parameters of the fitting procedure, which makes use of the input photometric catalog including the information of photometric redshift.

To avoid degeneracies between similarly probable solutions, we tailored the set of templates considering only the two populations we expected to be included in our selection. We used two evolving templates with solar metallicity and Chabrier IMF: an exponentially declining star formation history built with BC03 SPS, with timescale of $0.1\,$Gyr, representing a quiescent galaxy, and a template with constant star formation including emission lines, built with \emph{fsps} \citep{Conroy09,Conroy10}, appropriate for star-forming galaxies. Each template contains $221$ spectra for evolving ages from $t = 0$ to $t = 20$ Gyr. No formation redshift was imposed but only ages smaller than the age of the universe at the photometric redshift were considered. Other population synthesis models are available from the literature (e.g., \citealt{Maraston05}, hereafter M05); different choices could lead to variation in the stellar masses of $0.10-0.15$ dex depending on the considered SPS model (e.g., \citealt{Walcher11,Ilbert13}). For consistency and continuity with previous works (e.g., \citealt{Laigle16,Ilbert13, Davidzon17,Muzzin13}), BC03+fsps models were adopted as reference templates here. 

Moreover, we performed several SED-fitting runs using different extinction laws: in particular, those of \citet{Calzetti00}, which are characteristic of starburst galaxies, \citet{Fitzpatrick86} for the Large Magellanic Cloud, and \citet{Seaton79} for the Milky Way (see Sect.~\ref{sec_extlaws} for the expected differences in the color selection). We chose as our reference the results obtained using the law of \citet{Calzetti00}. We set the range of extinction between $A_V=0.0$ and $A_V=5.0$ both for the $\tau$ model and for star forming galaxies. Although high values of extinction are not expected in quiescent galaxies, we let the parameter assume all the values in the range to avoid any bias on the age, and consequently, on the stellar mass because of the well-known degeneracy among these parameters (e.g., \citealp{Papovich2001}).  As a result of making this choice, we expected to select galaxies that would be more consistent with being quiescent than dusty star-forming. 

The filters used for the fit are $u$, $B$, $V$, $r$, $i^+$, $z^{++}$,$Y$, $J$, $H$, $K_{\rm s}$, IRAC$[3.6]$ and IRAC$[4.5]$. The fluxes in the IRAC$[5.8]$ and IRAC$[8.0]$ bands have not been included in the final fit because of the shallower depth and worse PSF (point spread function) of these bands, which makes them much less suitable for constraining the SED shape. The filters used in the fit sample the optical/NIR wavelengths well enough even at the high redshifts considered in this work, so that the stellar mass can still be reliably determined \citep[see e.g., ][]{Taylor11,Pacifici12,Mitchell13,Conroy13}.  To avoid the problem of dealing with non-detections that should require a statistical treatment as in \citet{Sawicki2012}, we directly used fluxes (corrected to total) with their face values and errors. In this way, the $\chi^2$ can be estimated also for very small or negative values of the measured fluxes, as explained in \citet{Laigle16}, and the number of data points used to constrain the fit is always $\ge 9$.

Moreover, the intergalactic medium average opacity of intervening systems along the line of sight of high redshift objects has been taken into account following \citet{Madau95}.

\subsection{Results of SED fitting procedure}

\begin{figure*}[ht]
\centering
 \includegraphics[width=18cm]{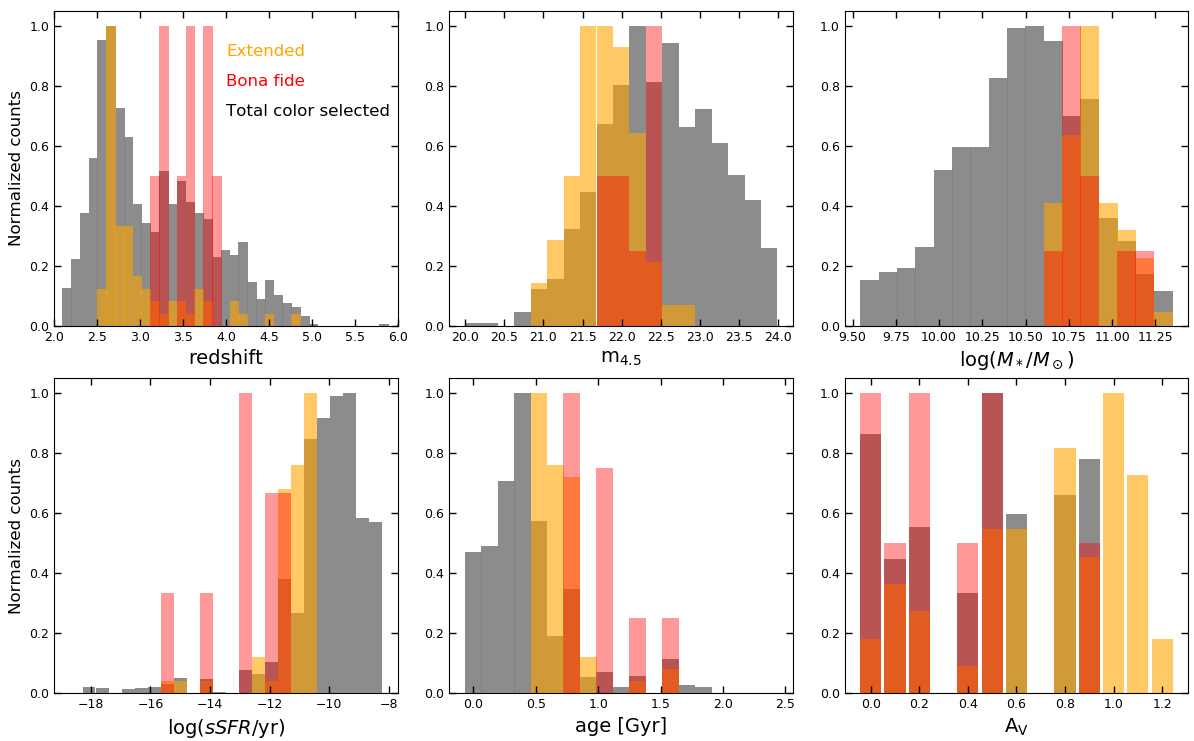}
 \caption{Distribution of physical properties of the color selected sample (gray) of the extended (orange) and bona fide (red) sub-samples.}
 \label{prop}
\end{figure*}

We divided our color-selected sample in two classes (the extended and the bona fide samples), following the results of the SED fitting. 

The extended sample is defined to have  $z_{\mathrm{phot}}\ge 2.5$, $\log (M_*/M_\odot)\ge 10.6$, $\log ({\rm sSFR\, [\,yr^{-1}]}) \le -10.5$ and the reduced $\chi^2$ of the best fit with the model of quiescent galaxies smaller than the one using the model with constant star formation ($\chi^2_{\mathrm{q}}<\chi^2_{\mathrm{sf}}$).

The bona fide sample, characterized by more restrictive criteria, i.e.,  $z_{\mathrm{phot}}\ge 3.0$, $\log (M_*/M_\odot)\ge 10.6$, $\log ({\rm sSFR\, [\,yr^{-1}]}) \le -11$, $\chi^2_{\mathrm{sf}}-\chi^2_{\mathrm{q}}>2$, and $\chi^2_{\mathrm{q}}<2$.

In particular, we choose the stellar mass cut at $\log (M_*/M_\odot)\ge 10.6$ in order to guarantee the completeness of our samples and facilitate the comparison with results and mass functions that have previously been published in the literature. 
We present the results of all the selections in Table~\ref{table:2}. The total number of objects in the bona fide sample is nine. The extended sample is made up of $128$ objects.  By using the different extinction laws mentioned in the previous section (i.e., \citealt{Fitzpatrick86,Seaton79}), we obtain almost identical results.  In further detail, we find that eight objects of the bona fide sample were selected independently from the extinction law adopted in the fit, while the object left out from our reference sample is the one with the largest amount of dust ($A_V=0.9$) in Fig.~\ref{prop}. This object is fitted with a value of $A_V=1.0$ by Fitzpatrick and Seaton extinction laws, but is excluded from the bona fide sample due to the higher sSFR ($\log({\rm sSFR\,[\,yr^{-1}]})\approx -10.9$ both for Fitzpatrck and Seaton laws). Also, the extended sample shows little variation compared to our reference results derived using \citet{Calzetti00} extinction law: we select $131$ objects assuming Fitzpatrick’s law, and $137$ with Seaton’s law.
In the color-color diagrams illustrated in Fig.~\ref{cand}, together with the parent sample in gray and candidates selected in Sect.~\ref{candidates} in black, we show the bona fide and extended samples with red and orange points, respectively. Red and orange points below the selection regions correspond to objects that are non-detected in one filter ($H$ or $J$): they have been fully considered in the samples as their color is consistent with the selection criteria when considering it as a lower limit.

\begin{figure*}
{\includegraphics[width=0.49\textwidth]{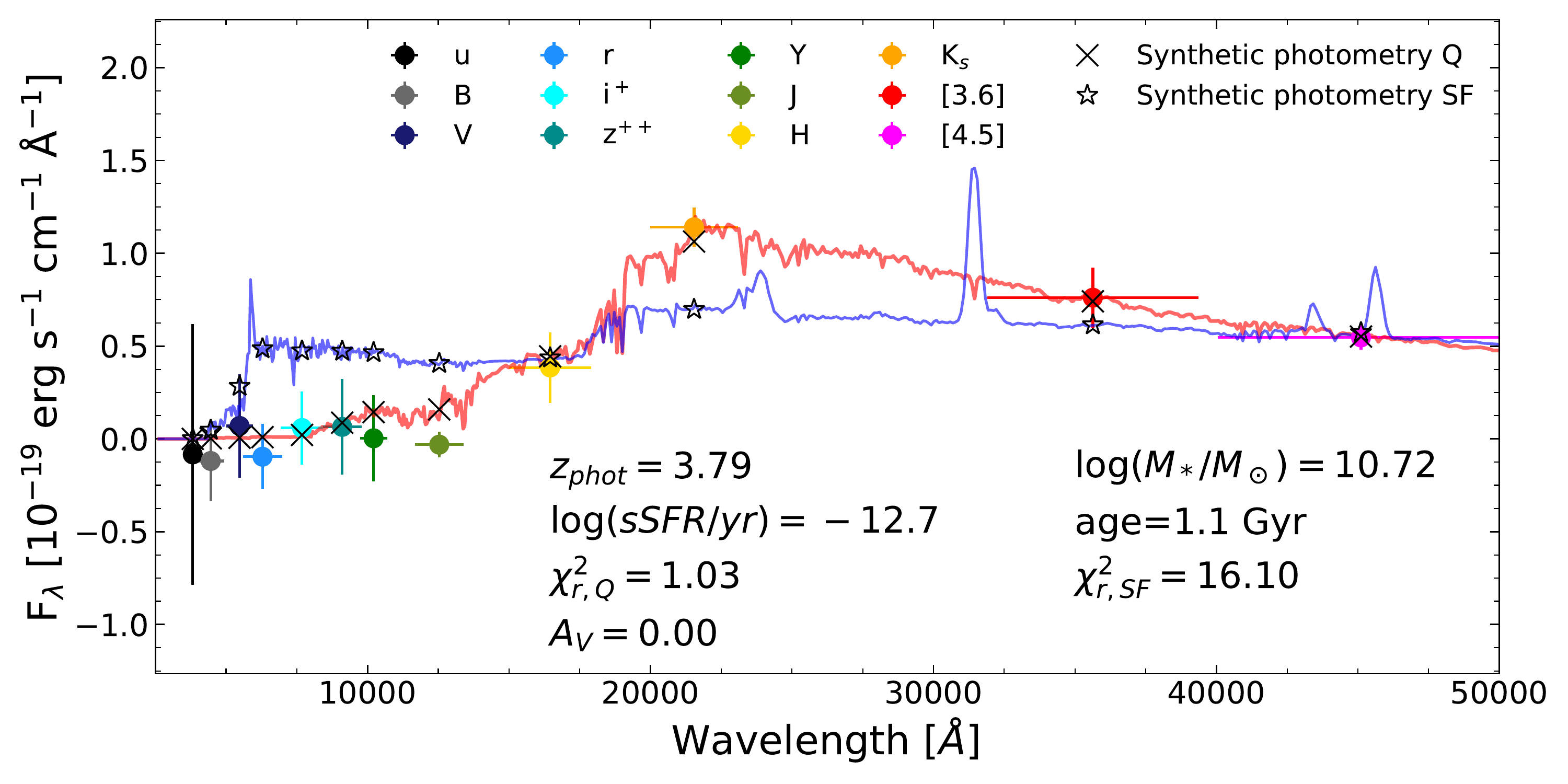}
\includegraphics[width=0.49\textwidth]{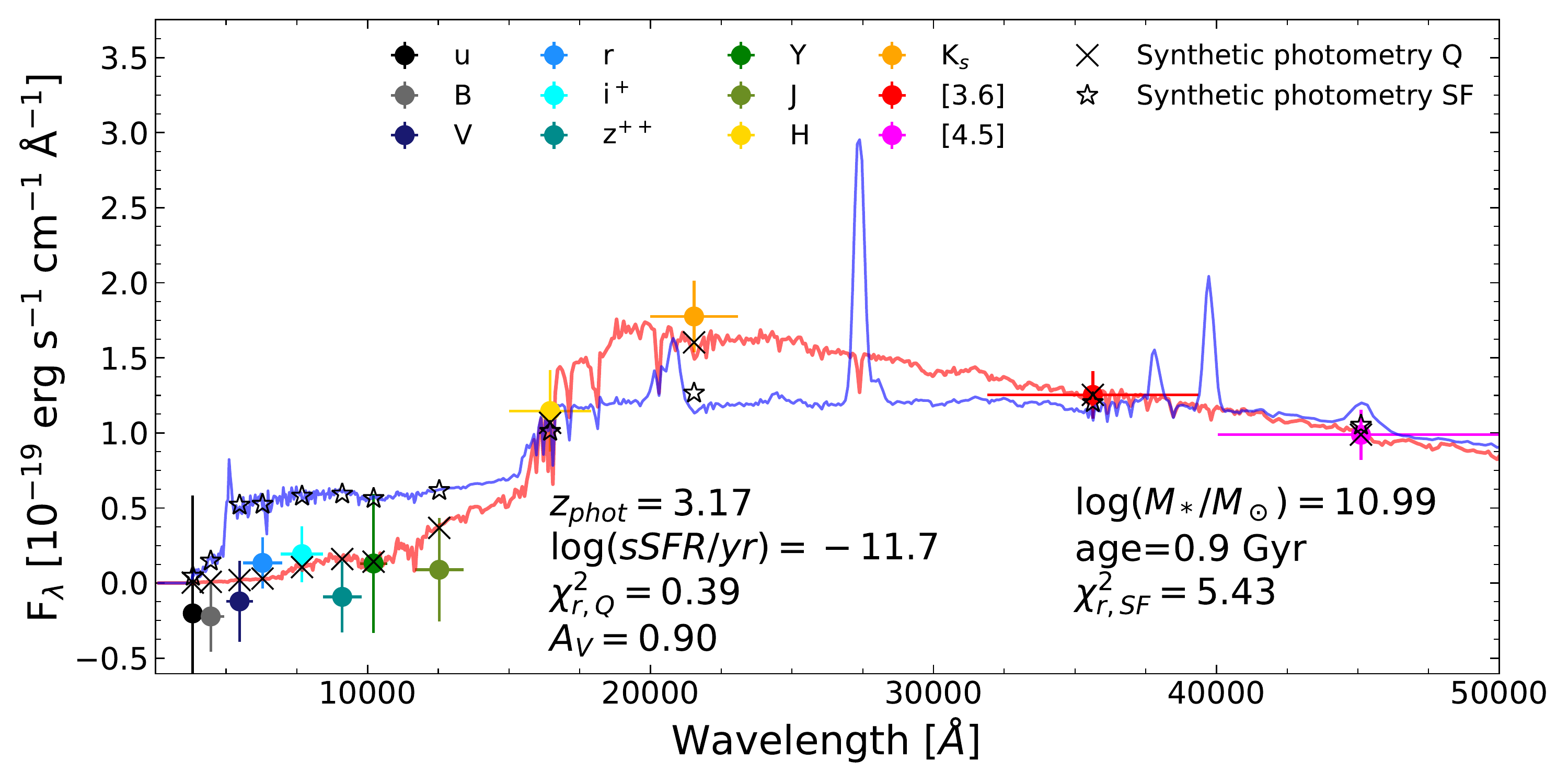}}\\
{\includegraphics[width=0.49\textwidth]{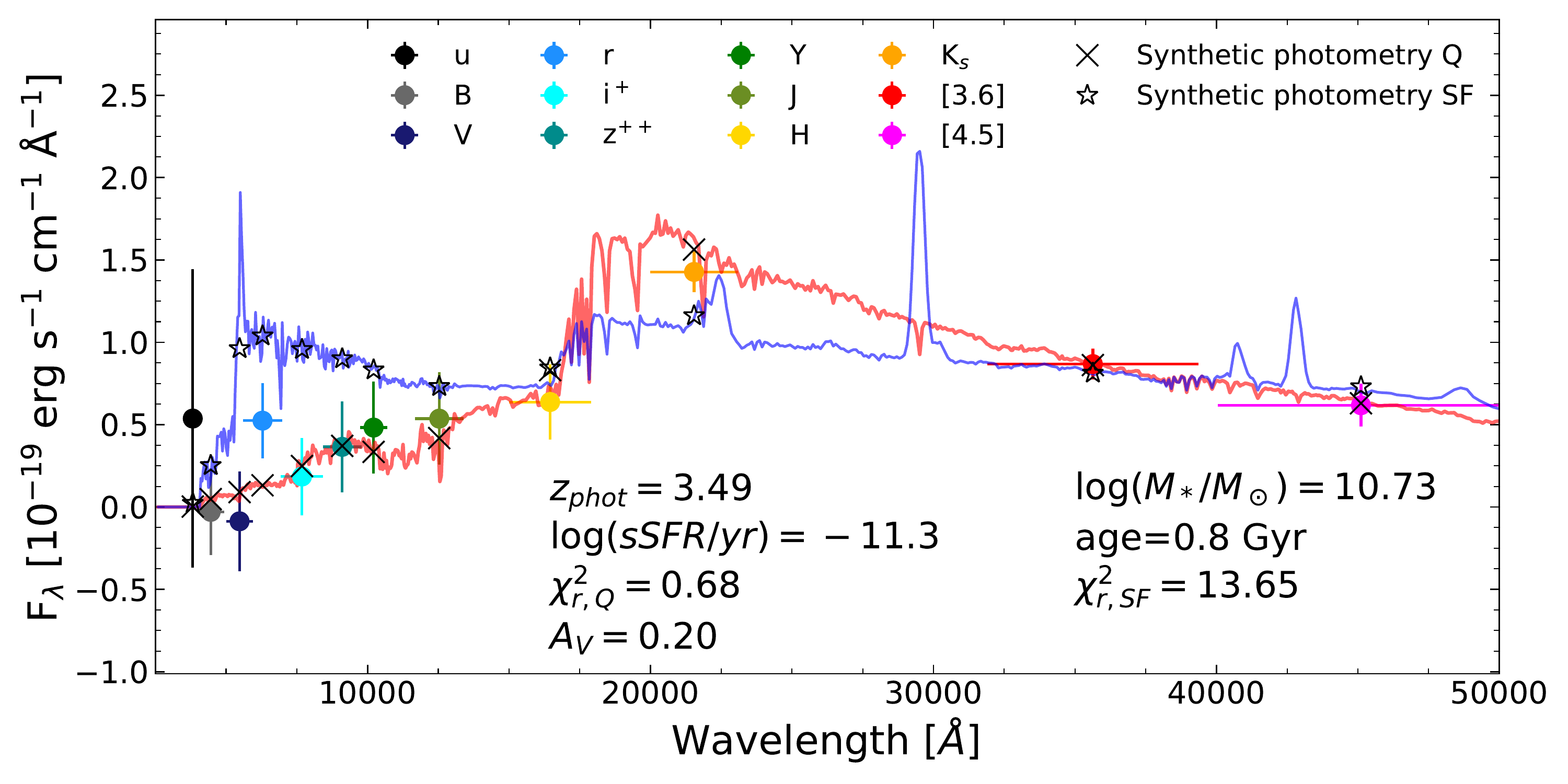}
\includegraphics[width=0.49\textwidth]{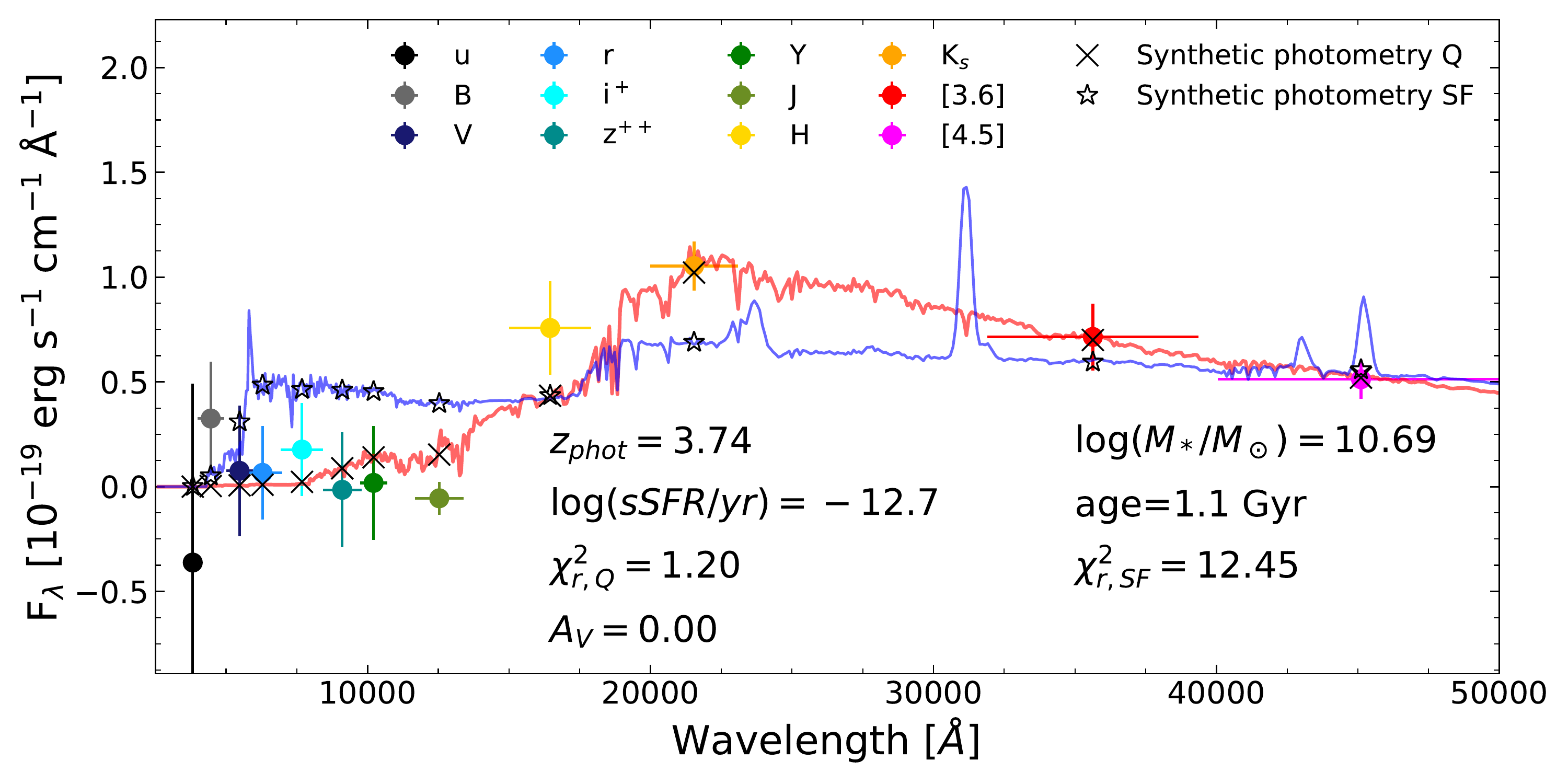}}\\
{\includegraphics[width=0.49\textwidth]{bonafide_475092.pdf}
\includegraphics[width=0.49\textwidth]{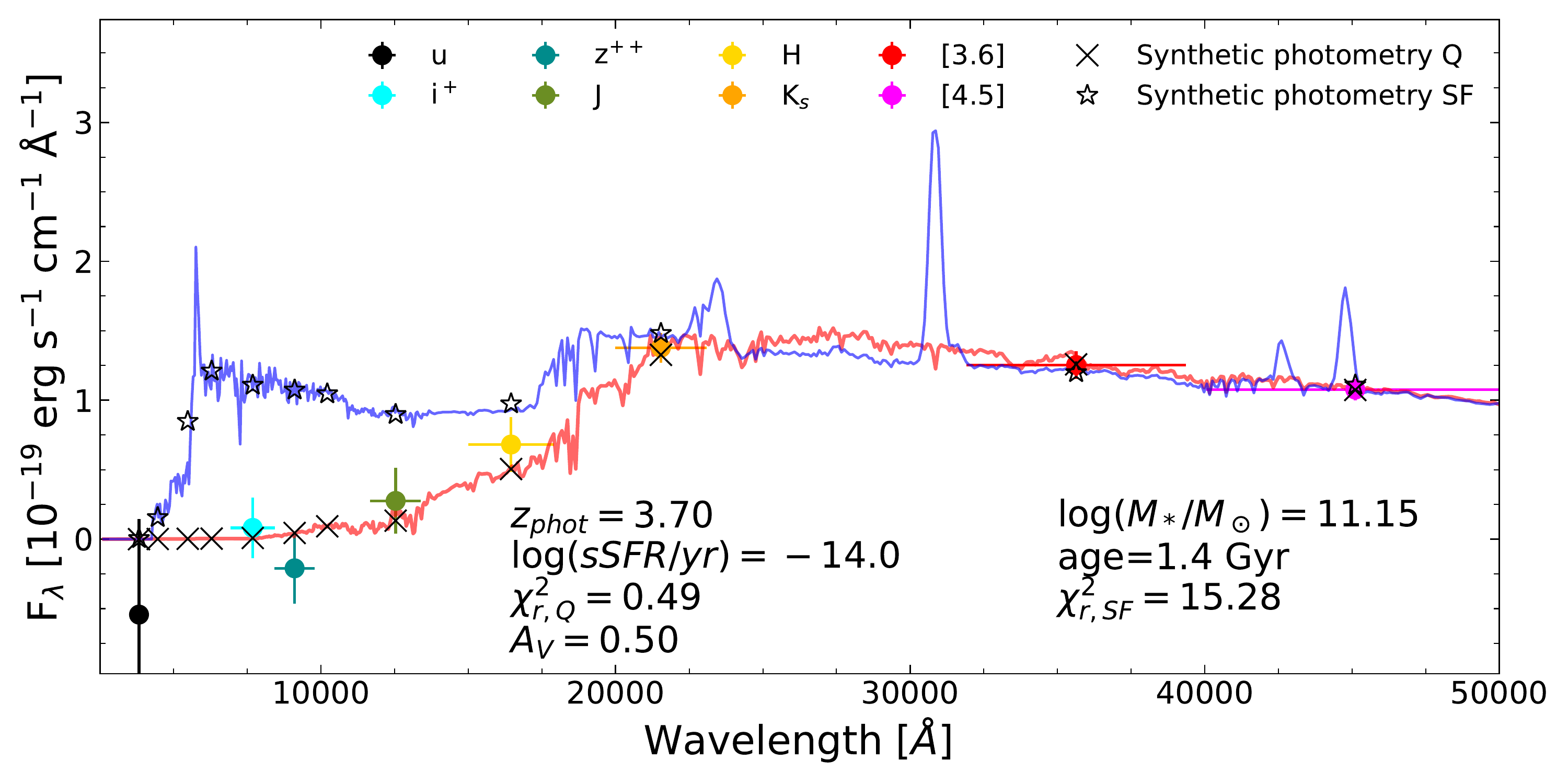}}\\
{\includegraphics[width=0.49\textwidth]{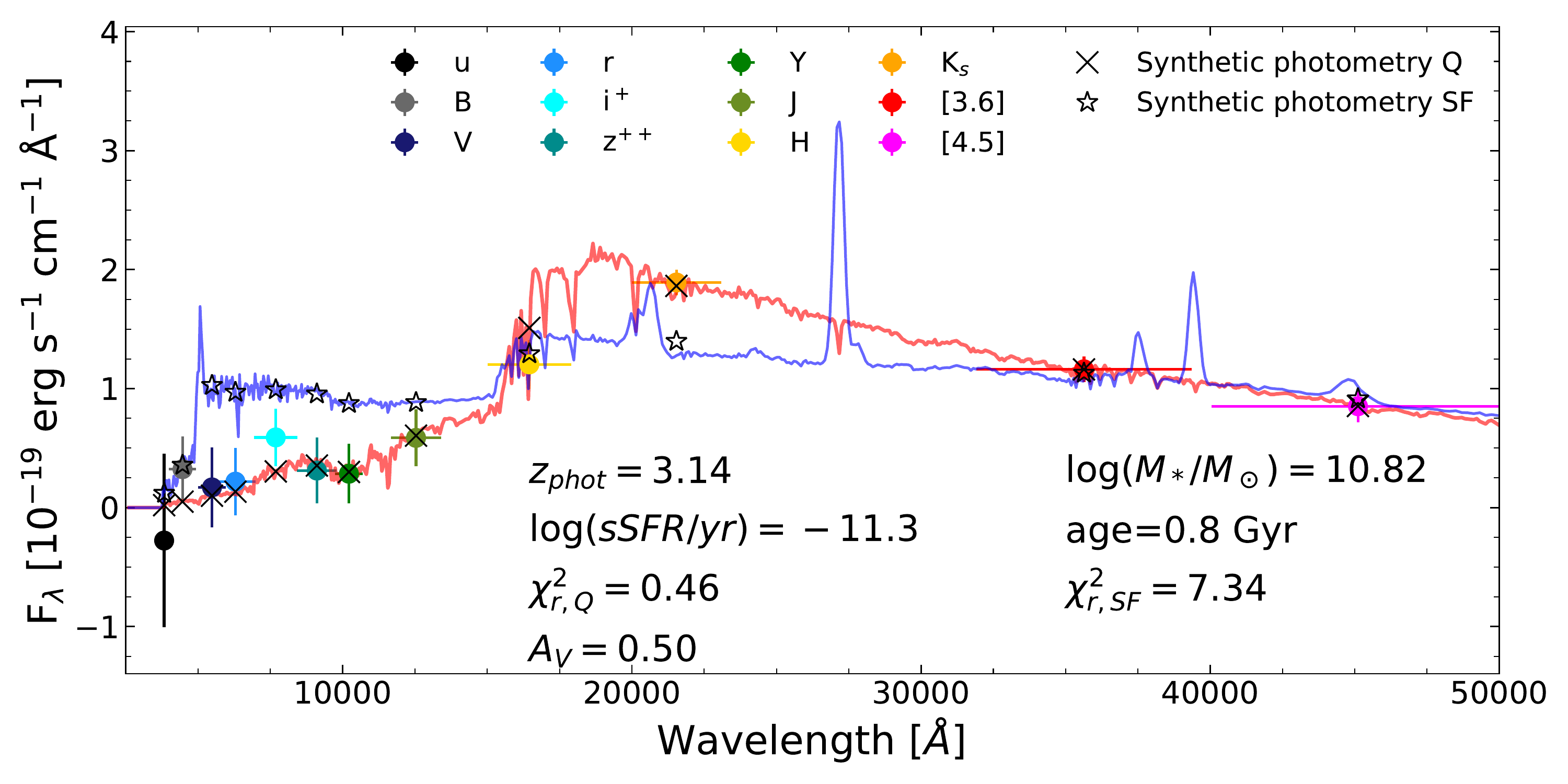}
\includegraphics[width=0.49\textwidth]{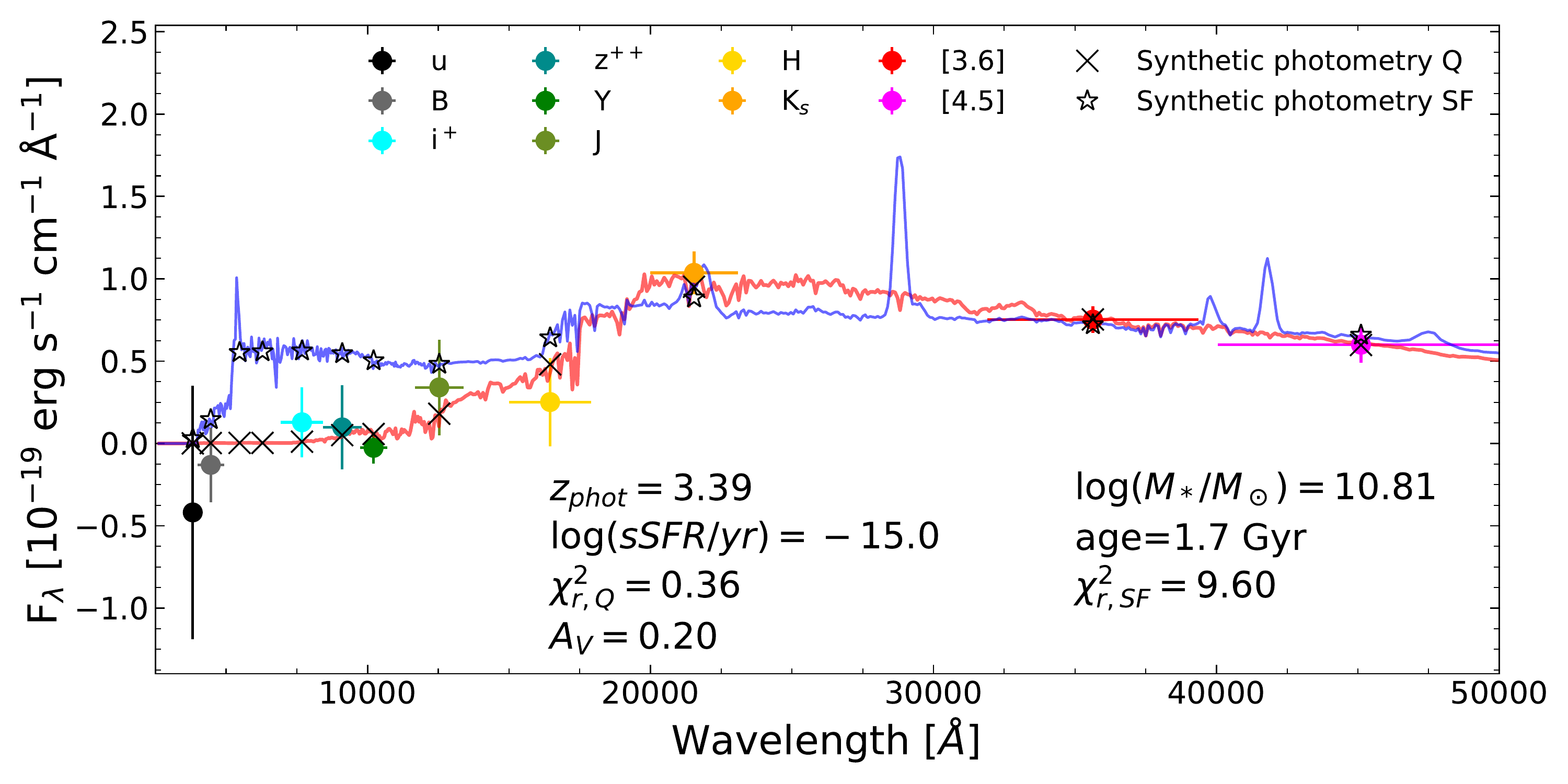}}\\
{\includegraphics[width=0.49\textwidth]{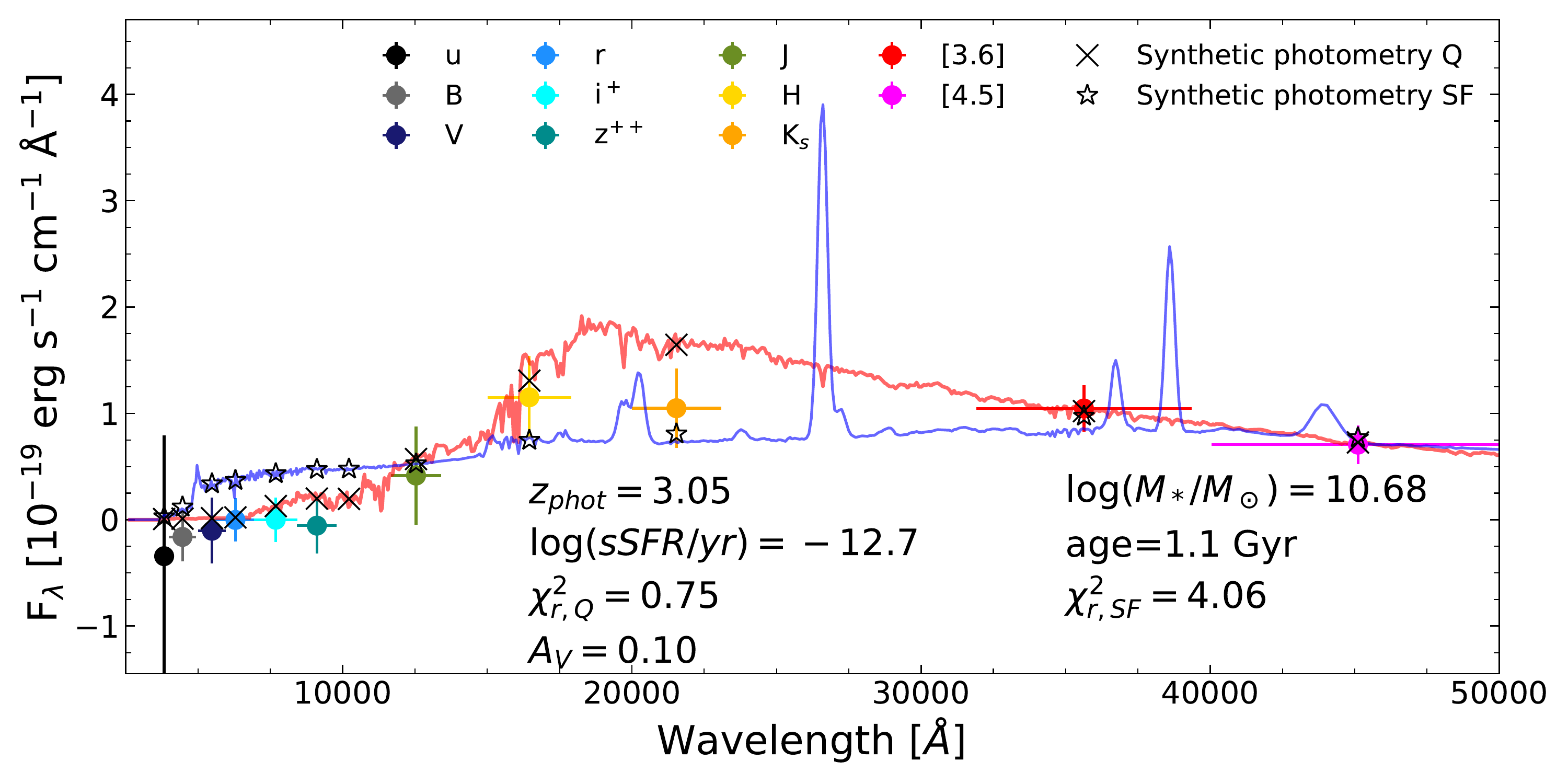}}

\caption{Best-fit models for all bona fide objects: colored points represent photometric points with error-bars, red lines are\ best-fit SEDs. The best fit using the star-forming template is shown with a blue line. Error bars in the wavelength scale refer to the width of the considered filter. Physical parameters evaluated through the SED fitting are shown in inserted labels. Black tilted crosses and stars show the synthetic photometry, evaluated by integrating the best-fit templates in the several bands, for the best-fit quiescent and star-forming templates respectively. All filters used in the fit are listed in inserted legends for each plot.}
\label{SEDex}
\end{figure*}    

We present in Fig.~\ref{SEDex}, the SED fitting results for all the galaxies in the \textit{bona fide} sample,  showing both the best-fit quiescent template, and the best-fit using only the star-forming template. 
The star-forming model at the same redshift is not able to reproduce the break strength located between the bands $H$ and $K_{\rm s}$, nor the faint fluxes at blue/near UV rest frame wavelengths. Similar results for the fit with the two classes of SEDs are valid for all the other objects: on one side, the inclusion of emission lines is not sufficient to mimic the D4000 break, while on the other, the fluxes blue-ward of the break are too high despite the large values allowed for dust extinction.

Figure~\ref{prop} summarizes the main physical properties of the extended and bona fide samples derived through the SED fitting analysis, taking as a reference fit the one using Calzetti's extinction law. In particular, the bona fide sample is shown in red while the extended sample is in orange. The properties shown are the photometric redshift and the observed magnitude at $4.5\,\mu$m from the COSMOS2015 catalog,  the stellar mass, the sSFR, the age, and the dust extinction from the SED fitting analysis. Many of the extended sample objects are located at $2.5 \lesssim z\lesssim 3.0$, whereas for the bona fide sample, the median redshift is $z\approx 3.38$. Compared to the total sample selected in Sect.~\ref{candidates} (shown in gray), the distributions of $m_{[4.5]}$ peak at brighter magnitudes. The other panels of the figure show how the \textit{bona fide} and \textit{extended} samples represent the most massive and quiescent galaxies chosen through color selections. We notice that three galaxies of the \textit{bona fide} sample have best fit with $0.5 \leq A_V \leq 0.9$, which are extinction values approaching those of Lyman-break galaxies at the same redshifts \citep{Shapley05}. However, recent works \citep{Gobat18,Martis19} claim that quiescent galaxies at high redshifts ($1<z<4$) can contain at least two orders of magnitude more dust at a fixed stellar mass compared with local early-type galaxies (ETGs). They found that these dusty ($A_V \ge 1.0$) high-$z$ quiescent galaxies can comprise up to $\sim 20-25\%$ of the population of quiescent galaxies, which is consistent with our findings. 
Moreover, we stress that the results we find are dependent on the population synthesis model we adopt. In fact, if we adopt a different SPS model (e.g., \citealt{Maraston05}), the best-fit $A_V$ value may change, along with other physical parameters degenerate with dust extinction, such as the age of the galaxy. We explore this possibility, using the SPS model of \citealt{Maraston05}  in the following Section.

Out of our $128$ quiescent candidates (extended sample) we find that $98$ of them are also classified as quiescent in the COSMOS2015 catalog, where quiescent galaxies are identified using the locations of galaxies in the color–color plane ${\rm NUV}-r/r-J$ \citep{Williams09}.

\begin{table*}[ht]
\caption{All objects selected through color selections and results of BC03 SED fitting, that is, both objects detected in all bands of interest (i.e., $J$,$H$,$K_{\rm s}$,$[3.6]$,$[4.5]$) and also objects with one or more non-detections in bands of interest.} 

\label{table:2}      
\centering  
\begin{tabular}{c|c c c c c }     
\hline\hline 
Number of objects & Color selection &  Number of objects & Number of objects &  Number of objects & Number of objects\\ 
 in parent sample& & color selected& color selected& in \textit{extended}& in \textit{bona fide}\\
& &with $m_{4.5}\leq 24$&with $M_*> 10^{10.6}\,M_\odot$&sample&sample\\ 
& &&$m_{4.5}\leq 24$ \& $z_{\rm phot}\geq 2.5$&&\\
\hline

212897 & $HK_{\rm s}$[3.6]       & 768  & 160  & 43  & 7\\
       & $JK_{\rm s}$[3.6] [4.5] & 462  & 263 & 95  &  9\\
\hline                  
\end{tabular}
\end{table*}

\subsubsection{M05 vs BC03 models}
\label{BC03vsM05}

We derive the best-fit physical parameters also using M05 models \citep{Maraston05} to evaluate the robustness of our selections. 
The two models (i.e., BC03 and M05) differ in the treatment of the thermally-pulsing asymptotic giant branch (TP-AGB) phase. The contribution of the TP-AGB stars to the integrated light of a synthetic stellar population critically depends on what is adopted for the stellar mass loss during this phase. The higher the mass loss, the sooner the star loses its envelope and the sooner the TP-AGB phase is terminated. In M05 the TP-AGB phase contribution is much higher than in BC03 models. The result is that the M05 models are brighter and redder than the BC03 models for ages between $\sim 0.2$ and $\sim 2$ Gyr at $\lambda >2-2.5\,\mu$m  \citep{Maraston06}. 
This implies that the M05 models give, in general, younger ages and lower stellar masses compared to BC03 models. At older ages, this tendency is reversed.

In order to perform a fair comparison we apply the same criteria used to define the bona fide and extended samples described at the beginning of the section.  While we adopt for BC03 models a Chabrier IMF for consistency and continuity with other works in the literature  (e.g., \citealt{Ilbert13,Laigle16,Muzzin13,Davidzon17}), for M05 models, a Chabrier IMF is not available, so we adopt a Kroupa IMF \citep{Kroupa01} and apply a statistical offset to stellar masses, such as $\log M_{*,\rm Chabrier} = \log M_{*,\rm Kroupa}-0.04$, to take into account the different IMF.  All other parameters are consistent between the two models. Moreover, for the M05 SED fitting run we also adopted the star-forming template that includes emission lines built with \emph{fsps}.

Using M05 models to fit the data, $20\%$ fewer objects are included in the bona fide sample with respect to BC03 results (i.e., eight objects are selected in the M05 bona fide sample), while $\approx 33\%$ fewer objects are included in the extended sample. The case of having different numbers of objects is expected since M05 models predict for ages between $\sim 0.2$ and $\sim 2$\,Gyr (i.e., the typical ages for galaxies at  $z\gtrsim 3$) a higher flux per unit mass with respect to BC03 models \citep{Maraston06} at $\lambda\gtrsim 2\,\mu$m, due to a higher contribution from stars in the TP-AGB phase. This reflects in a lower normalization needed for the theoretical SED to fit the observed data, and ultimately to a lower mass estimate. As a consequence, fewer objects satisfy the $\log (M_*/M_\odot)\ge 10.6$ condition.

To have a fair comparison between the number of objects selected with BC03 and M05 models, we compute the median offset of the mass estimation between BC03 and M05 models of the bona fide and extended samples. We find that the median value of  $\log(M_{*,\mathrm{BC03}}/M_{*,\mathrm{M05}})$ is $\approx 0.24$ when considering the extended sample, which is larger than the value of $0.14$ used by \citet{Henriques15}.  However, the value adopted by \citet{Henriques15} was derived by \citet{Dominguez11} and \citet{Pozzetti10} for a mix of different galaxy populations, while the difference is expected to be larger for objects that quenched their star formation. Assuming a lower mass cut of $\log (M_*/M_\odot)\ge 10.6-0.24=10.36$, the total number of objects in bona fide and extended samples is more in agreement with the results obtained using BC03 models, which are, respectively, $8\%$ smaller and $33\%$ bigger than the BC03 results. 

We also investigated the dependence of the derived age on the adopted model. If we leave the dust extinction free to change in the process of SED-fitting, then similar ages are obtained either with M05 or BC03 models (both for the extended and bona fide samples). However, the BC03 model predicts a mean extinction value of $0.81$ mag for the extended sample and $\left \langle A_V\right\rangle=0.31$ for the bona fide, while M05 models predict a mean extinction of $0.17$ and $0.09$ mag for the extended and bona fide samples, respectively. If, instead, we do not allow for dust extinction (i.e., we fix $A_V=0$), the mean values of the ages predicted by the two models differ, with M05 models predicting $\approx 0.4$\,Gyr younger ages with respect to BC03 models. These results are in agreement with the work presented in \citet{Maraston06}, where similar conclusions are found for a sample of seven passive galaxies at $z>2$ when dust extinction is forced to assume only very small values.

There is no consensus, yet, in the literature about which of the two models better represents high redshift galaxies, in other words, about the relative importance of the TP-AGB phase in high redshift objects. In fact, while, for example, \citet{Maraston06} finds a better fit of high-$z$ galaxies using M05 models, other studies, such as that of \citet{Kriek}, find better fits using BC03 models. In this paper, we do not find any strong evidence in favor or against any of the two considered SPS models.  In the following of this paper, we will use BC03 results to ease the comparison with previous work (e.g., \citealt{Davidzon17}, \citealt{Ilbert13}, \citealt{Muzzin13}).


\section{Number and mass densities}
\label{numdens}

\begin{table*}[ht]
\caption{Bona fide sample as selected by means of SED fitting with BC03 models: number and mass densities in different redshift bins. Errors include Poissonian errors, cosmic variance, and photometric redshift errors. Note: no objects in the bona fide sample are located at $z\ge 4$.}          
\label{table:3}      
\centering          
\begin{tabular}{cccccc}     
\hline\hline
$\Delta z$ & \small Number&$z$& \small $\log (M_*/M_\odot)$&$\log(\rho_N)$&$\log(\rho_*)$\\  
  &$bona\,fide$&median&median&($\mathrm{Mpc}^{-3}$)&($M_\odot \,\mathrm{Mpc}^{-3}$)\\ \hline
$3.0\le z<3.5$ & $5$ & $3.17$ & $10.81$ & $-6.044^{+0.173}_{-0.293}$ & $4.614^{+0.172}_{-0.290}$\\
$3.5\le z<4.0$ & $4$ & $3.72$ & $10.70$ & $-6.283^{+0.191}_{-0.351}$ & $4.599^{+0.157}_{-0.249}$\\
\hline                 
\end{tabular}
\end{table*}

\begin{table*}[ht]
\caption{Extended sample as selected by means of  SED fitting with BC03 models: number and mass densities in different redshift bins. Errors include Poissonian errors, cosmic variance, and photometric redshift errors added in quadrature.}             
\label{table:4}      
\centering          
\begin{tabular}{cccccc}    
\hline\hline
$\Delta z$ &  \small Number&$z$& \small $\log (M_*/M_\odot)$&$\log(\rho_N)$&$\log(\rho_*)$\\  
  &$extended$&median&median&($\mathrm{Mpc}^{-3}$)&($M_\odot \,\mathrm{Mpc}^{-3}$)\\ \hline
$2.5\le z<3.0$ & $98$ & $2.60$ & $10.81$ & $-4.924^{+0.080}_{-0.098}$ & $5.949^{+0.077}_{-0.095}$\\
$3.0\le z<3.5$ & $14$ & $3.30$ & $10.83$ & $ -5.757^{+0.126}_{-0.178}$ & $5.123^{+0.119}_{-0.164}$\\
$3.5\le z<4.0$ & $9$ & $3.62$ & $10.76$ & $-5.930^{+0.148}_{-0.228}$ & $4.843^{+0.147}_{-0.224}$\\
$z \ge 4.0$    & $7$ & $4.03$ & $10.82$ & $-6.018^{+0.164}_{-0.268}$ & $4.912^{+0.144}_{-0.218}$\\
\hline      
\end{tabular}
\end{table*}

\begin{figure*}[ht]
\centering
\includegraphics[width=18cm]{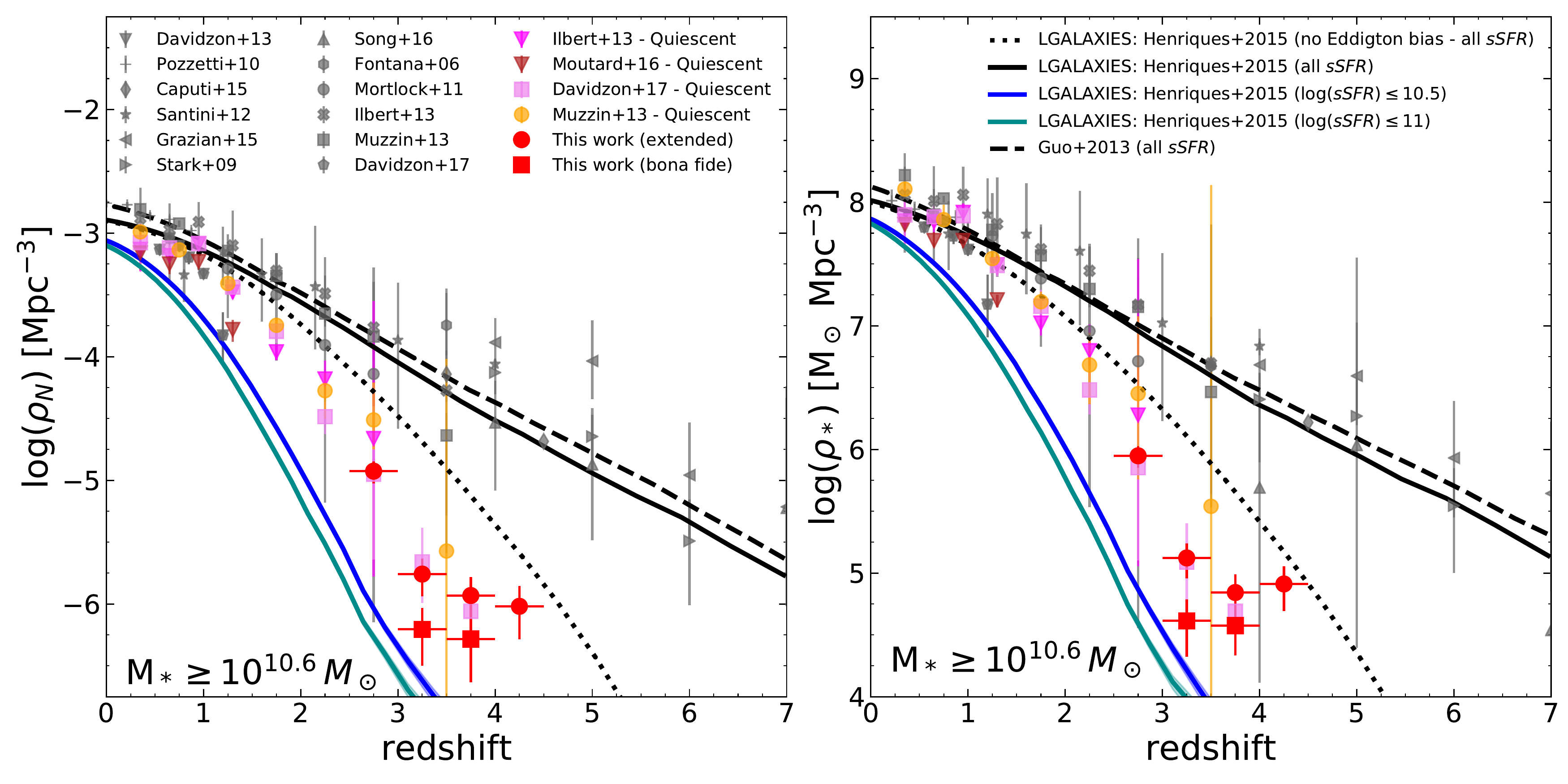}
\caption{Number and stellar mass densities of galaxies with $M\ge 10^{10.6}\,M_\odot$ as function of redshift for total population of galaxies (in gray) or quiescent galaxies (colored points) in literature, compared to those obtained in this work. Our results are shown as red diamonds and squares for the BC03 extended and bona fide samples, respectively, with error bars representing the Poissonian errors, cosmic variance, and photometric redshift errors added in quadrature. The forecasts of two semi-analytic models are also shown: the black dashed and continuous lines represent the \citet{Guo13} and the H15 models, respectively, estimated for the entire population of galaxies and convolved for the Eddington bias.  The black dotted line represents the H15 model for the entire population not convolved for the Eddington bias. Blue and cyan shaded regions represent the H15 model with Poissonian error for the selected quiescent populations characterized by $\log ({\rm sSFR\, [\,yr^{-1}]}) \le-10.5$ and $\le-11$}, respectively. 
\label{dens}
\end{figure*}

We use the results obtained in the previous section to estimate the number and stellar mass densities of quiescent galaxies at $z>2.5$ and compare them to observations and state-of-the-art semi-analytical models.
In Tables~\ref{table:3} and \ref{table:4}, the number and mass densities of the bona fide and extended samples are presented in redshift bins and shown in Fig.~\ref{dens}. To evaluate the number and mass densities, we consider only massive galaxies with $M\ge 10^{10.6}\,M_\odot$, both for the extended sample (red diamonds) and for the  bona fide sample (red squares).

The error bars we show in Fig.~\ref{dens} have been evaluated by adding in quadrature Gaussian errors for the object counting  (we approximated Poissonian statistic with the Gaussian one given that for large counts, the two statistics match), cosmic variance, and the scatter between different redshift bins due to photometric redshift errors. To take into account cosmic variance, we follow the prescriptions of \citet{Moster11}. In particular, we choose the field size to be consistent with the COSMOS field we are considering, we impose the same mass cut we adopt (i.e., $\log(M_*/M_\odot)\ge 10.6$) and for each redshift bin we choose the median redshift of the objects grouped in that bin. We find that, for our mass range, the cosmic variance ranges from a value of $\sigma_{\rm CV}=0.15$ at $2.5\leq z<3.0$ to $\sigma_{\rm CV}=0.25$ from at $z\ge 4.0$.
To take into account the scatter of objects between different redshift bins due to photometric errors, we checked how much the number of objects in the adopted bins can change by applying a noise of $\sigma_z \sim 0.03(1+z)$ to the redshift distribution of a sample selected with $m_{4.5}<24$. We used the COSMOS catalog itself and a lightcone from H15 model to estimate the variation in the number of objects in high redshift bins, and we find that it is always less than $\sim 7\%$. We adopted an additional error of $10\%$ on our measurements of number and mass densities. We notice that the assumption on the scatter can be optimistic given the lack of spectroscopic control samples for this class of objects, but even assuming a scatter as large as $\sigma_z \sim 0.10(1+z)$ , we expect a contamination of $11\%$ in the wide bin $3.0\le z<4.5$ adopted in Sect.~\ref{massfct}, and a maximum contamination of $\sim 20\%$ at $3.5\le z<4.0$ and $\lesssim 50\%$ at $4.0\le z<4.5$. 

We find the following results.
For the extended sample:\ we find a decrease by a factor of $\sim 14\,^{+3}_{-7}$ in the number of quiescent object from the lowest redshift bin (i.e., $2.5\leq z<3.0$) to the highest one (i.e., $z\ge 4.0$), and the same decrease is found by considering the mass densities between the same redshift intervals. In addition, we find a decrease by a factor $\sim 66\,^{+28}_{-27}$ and 
$\sim 100\,\pm ^{+29}_{-35}$ in the number density and mass density, respectively, between \citet{Davidzon17} data for the quiescent population at $0.2<z<0.5$ and our estimate in the redshift bin $2.5\leq z<3.0$.  Between the same redshift bins, considering the results of \citet{Moutard16B}, which are evaluated on the VIPERS multi-lambda catalog (on $22$\,deg$^2$) and are, therefore, more statistically significant and less subject to cosmic variance with respect to results on COSMOS, we find a decrease by a factor $\sim 57\,^{+14}_{-19}$ and 
$\sim 78\,\pm ^{+15}_{-19}$ in the number density and mass density, respectively.
Concerning the bona fide sample: between the lowest redshift bin (i.e., $3.0\leq z<3.5$) to the highest  (i.e., $3.5\leq z<4.0$) we find an evolution by a factor of $\sim 1.7$ in the number density and almost no evolution in the mass densities. 

The errors on the estimated factors have been evaluated by propagating the errors of our measurements. When considering the \citet{Davidzon17} data, the errors on their measurements have also been considered and propagated.
Interestingly, the $10\%$ uncertainty we add to our error budget in order to take photometric redshift errors into account, is similar to what \citet{Ilbert13} quoted as error relative to the template fitting procedure, including photometric redshift error and stellar mass estimate uncertainties.

\subsection{Comparison with previous results}
\label{obs}

In Fig.~\ref{dens}, we compare the evolution of number and stellar mass densities with  results from the literature, specifically,\ \citet{Pozzetti10,Davidzon17,Santini12,Ilbert13,Davidzon13,Muzzin13,Mortlock11,Fontana06,Caputi15,Stark09,Song16,Grazian15}. The datapoints have been rescaled where necessary to the cosmology we adopted (see Sect.~\ref{intro}) and to a Chabrier IMF. We consider only massive galaxies with $M\ge10^{10.6}\,M_\odot$ both for the full sample of galaxies (gray points) and for the sub-sample of quiescent galaxies (colored points). The literature we used featured data which focused on quiescent galaxies (\citealt{Davidzon17,Ilbert13,Muzzin13,Moutard16B}).  All of them, like the present work, collect a galaxy sample where photometric redshifts and stellar masses are derived via SED fitting.  The classification of quiescent galaxies is based on the sSFR in the present work, whereas it is based on the ${\rm NUV}rJ$ diagram for \citet{Davidzon17}, \citet{Moutard16B} and \citet{Ilbert13}, and on the $UVJ$ diagram for \citet{Muzzin13}. We have been able to evaluate the number and mass densities of quiescent galaxies at the highest redshifts from the stellar mass functions that have been published. Data from \citet{Davidzon17} are estimated by means of the integral of their stellar mass function for the quiescent population, obtained from COSMOS2015 catalog. The same recipe has been applied to the stellar mass
function of quiescent galaxies of \citet{Muzzin13}  selected from a NIR-selected sample of galaxies out to $z \sim 4$, and to the mass functions of \citet{Moutard16B} at $z\le 1.5$ derived in a $22\,{\rm deg}^2$ field. The results of the present work are in broad agreement with the results obtained by \citet{Ilbert13}, \citet{Muzzin13}, and \cite{Davidzon17}.

\subsection{Comparison with semi-analytic models}
\label{H15}

We compared our results with the number and stellar mass densities obtained using semi-analytic models by \citet[][ hereafter G13]{Guo13} and \citet[][ hereafter H15]{Henriques15} for the total and the quiescent populations. Both models are based on the Munich galaxy formation model \citep{Kauffmann99,Springel01,Croton06,Delucia07,Guo11,Guo13,Henriques15} which has been implemented in the Millennium \citep{Springel05} simulation of dark matter in a box with comoving side of $500\,{\rm Mpc}\, h^{-1}$, and with cosmological parameters adopted from the Wilkinson Microwave Anisotropy Probe \citep{Komatsu06} and from the \citet{Planck14} for G13 and H15, respectively. 
We used the data from the Millennium database\footnote{http://gavo.mpa-garching.mpg.de/MyMillennium/ Help/databases/henriques2015a/ database}, selecting galaxies directly from the snapshots of the simulation.
Before applying the same selection criteria adopted for observed galaxies, we convolved stellar masses with a Gaussian in $\log(M_*/M_\odot)$, with width increasing with redshift $\sigma_{\log M_*}=0.08(1+z)$, in order to account for the Eddington bias, as done in \citet{Henriques15}. We set the Hubble parameter as in our reference $\Lambda$CDM cosmology, and we selected galaxies with $\log(M_*/M_\odot)\geq 10.6$ at all sSFR for the total population shown in Fig.~\ref{dens}. For quiescent galaxies, we selected massive galaxies (i.e., $\log(M_*/M_\odot)\geq 10.6$) with two different cuts in $\log ({\rm sSFR\, [\,yr^{-1}]})$: one at $-10.5$ to match our extended sample definition and one at $-11$ to match the bona fide selection (shown in Fig.~\ref{dens} as blue and cyan lines respectively).
It can be argued that the different star formation histories assumed in our SED fit and in the models may bias the comparison between the values. However, it has been shown by \citet{Laigle19} that the instantaneous or short time-scale sSFR derived for a COSMOS-like photometric sample using exponentially declining SFHs reproduces the intrinsic values of a simulation with stochastic SFHs reasonably well.

In Fig.~\ref{dens}, we also show the effect of the correction for the Eddington bias for the total population in the H15 model: the convolution with the Eddington bias tends to increase the number counts of galaxies especially at high redshift, since more galaxies with small masses (which are more numerous than galaxies at high masses) are scattered upwards of the mass threshold than high mass galaxies scattered downwards of the same limit. 
Since the observed values are naturally the result of the convolution of intrinsic properties and observational errors, it is fundamental to take this bias into account in order to make a fair comparison between models and observations. Figure~\ref{dens} shows that the amount of the convolved error is what is needed to match the densities of massive observed galaxies at high redshift.

While the total number and mass densities by H15 and G13 are in good agreement with the literature data, Fig.~\ref{dens} shows that the same quantities for passive galaxies derived by H15 are underestimated when compared with results from the present work and from the literature at high redshift: even considering a large scattering in photometric redshifts as mentioned above, the decrease in number and mass densities is not sufficient to fully reconcile the results with the models, at least for the extended sample, as we discuss in Sect.~\ref{discussion}.

\section{Stellar mass functions}\label{massfct}

\begin{figure*}[ht]
\centering
\includegraphics[width=18cm]{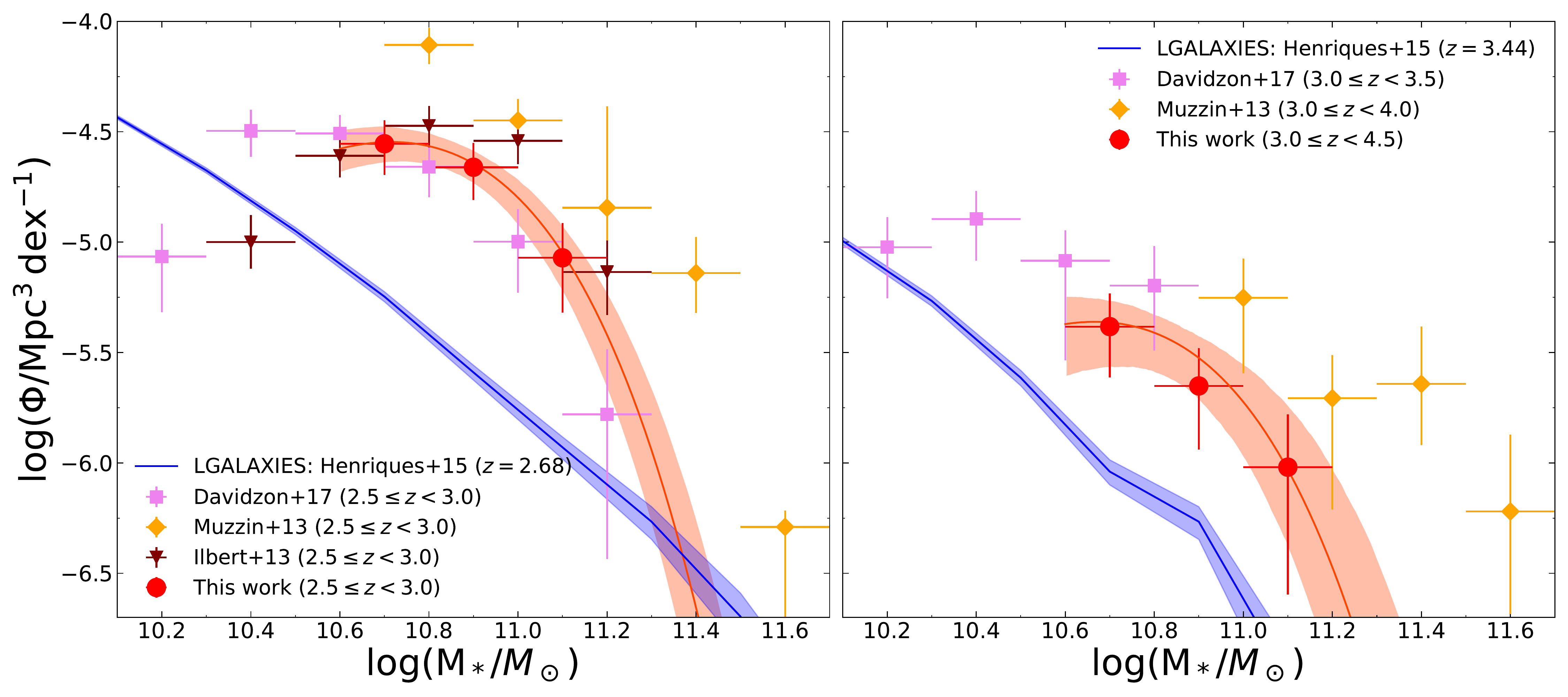}
\caption{Stellar mass function of \textit{extended} galaxies, in two redshift bins between $z = 2.5$ and $4.5$. In each panel, the data points are shown as red circles in bins of $\Delta\log M/M_\odot=0.1$. Error bars include Poisson noise, cosmic variance, and photometric redshift errors. The data points are fitted by a single Schechter function, shown by a red solid line, while the red shaded area is its $1\sigma$ uncertainty. Data points (with their error bars) from other works are also shown: violet squares are taken from \citet{Davidzon17}, in magenta triangles we plot the mass functions by \citet{Ilbert13}, and in orange circles the one by \citet{Muzzin13}. The blue lines shows the H15 model for quiescent galaxies at $z=2.68$ (left) and $z=3.44$ (right).}
\label{MF}
\end{figure*}

The stellar mass function (SMF) of galaxies has been studied extensively over the past years out to $z \sim 4 -5$, for both star-forming and quiescent galaxies (e.g., \citealt{Fontana06,Ilbert13,Davidzon17,Muzzin13,Grazian15}). In this section, we estimate the stellar mass function for quiescent galaxies at $z >2.5$ and compare it to other observations and models.

We estimate the stellar mass function of the \textit{extended} sample into two redshift bins: $2.5<z<3$ and $3.0<z<4.5$ with the $1/V_{\rm max}$ method (considering, for all the galaxies, the $V_{\rm max}$ equal to the volume defined by the redshift bin since we assume it to be complete for masses above $\sim 10^{10}\,M_\odot, $ as shown in Fig.~\ref{mlim}). We fit our data (3 mass bins for each redshift interval) with a Schechter function \citep{Schechter76}:
\begin{equation}\Phi (M) dM = \Phi^{*} \left( \frac{M}{M^*} \right)^{\alpha} \exp \left(- \frac{M}{M^*} \right) \frac{dM}{M^*}\end{equation}
using the software package \textit{emcee} \citep{Foreman12}, an MIT-licensed pure-python implementation of affine invariant Markov chain Monte Carlo (MCMC) ensemble sampler \citep{Goodman10}. Since we are probing only the massive end of the function, the slope of the power law part of the function is not constrained. We thus fix the parameter $\alpha=1.15$, following \citet{Davidzon17} and assuming the same slope they find at lower redshift. We report in Table~\ref{table:5} the Schechter parameters fitting the data points in the two redshift bins along with the $1\sigma$ errors.

\begin{table}[ht]
\caption{Schechter parameters of  best-fit stellar mass functions.}             
\label{table:5}      
\centering          
\begin{tabular}{cccc}    
\hline\hline
redshift & $\log(M^*)$&$\alpha$&$\Phi^*$\\  
 &[$M_\odot$]&(fixed)&[$10^{-5} \mathrm{Mpc}^{-3}$]\\
 \hline
$2.5<z<3.0$     & $10.38^{+0.05}_{-0.05}$ & $1.15$ & $10.21^{+1.80}_{-1.88}$\\
$3.0\leq z<4.5$ & $10.33^{+0.09}_{-0.12}$ & $1.15$ & $1.56^{+0.46}_{-0.57 }$\\
\hline                 
\end{tabular}
\end{table}

In Fig.~\ref{MF}, the data points along with the best-fit Schechter functions are shown. Error bars include the contribution of Poissonian errors, cosmic variance, and photometric redshift errors added in quadrature, while in the abscissa, they represent the mass bin $\Delta\log M/M_\odot=0.2$. For the cosmic variance, we once again used the prescriptions of  \citet{Moster11}, deriving for each bin the cosmic variance on the COSMOS field, at the median redshift of the objects grouped in each bin and in the mass range covered by the bin. To account for photometric redshift errors, we considered a $10\%$ error, as explained in Sect.\ref{numdens}.

We compared our SMF with the literature data in the same redshift bins  (rescaling masses to Chabrier IMF and $\Lambda$CDM cosmology when required).  We selected previous works that focused on quiescent galaxies (\citealt{Davidzon17,Ilbert13,Muzzin13}) and whose characteristics are explained in Sect.~\ref{obs}. In Fig.~\ref{MF}, we also show in blue the SMF for the H15 model (described in Sect.~\ref{H15}) that is derived in the snapshot located at $z=2.68$ for the redshift bin $2.5\leq z<3.0$ and at $z=3.44$ for the redshift bin $3.0\leq z<4.5$. We chose the snapshots located at redshifts that were the nearest to the median redshift of our observed galaxies in the two bins, which are $z=2.60$ and $z=3.35$.  We selected quiescent galaxies in the snapshots in the same way we selected the extended sample, that is, $\log(M_*/M_\odot)\geq 10.6$ and $\log ({\rm sSFR\, [\,yr^{-1}]}) \leq -10.5$, after convolving the masses with a Gaussian of width $0.08(1+z)$ in $\log(M_*)$. 
In Sect.~\ref{discussion}, we present an extensive discussion of our results and a comparison with results in the literature and semi-analytic models.

\section{Discussion}
\label{discussion}

\subsection{Comparison with previous observations} 

In Fig.~\ref{dens}, we compare our estimates of the number and mass densities with several results presented in the literature. The observed densities of the quiescent population are characterized by a rather small scatter: our results are in good agreement with \citet{Davidzon17} at all redshifts, while larger differences are visible with \citet{Muzzin13} and \citet{Ilbert13}, especially at $z<3$. This can be due to different input photometric datasets, SED fitting analysis performed on the data,  and the different criteria adopted to define a quiescent galaxy (see Sect.~\ref{obs}).

From the comparison of the observed SMFs in Fig.~\ref{MF}, it is evident that at $z>3.0$ \citet{Muzzin13} estimate a greater number of very massive ($> 10^{11} M_\odot $) quiescent galaxies than all the other works. 
The estimate by \citet{Ilbert13} is, instead, only slightly higher than ours at  $2.5<z<3$, while we obtain the best agreement with \citet{Davidzon17} in both redshift bins. Since all the SMFs are derived in the same field, differences cannot be simply ascribed to observations on different fields.

Based on Fig.~\ref{MF}, there is a noticeable evolution between the two redshift bins. This effect can be due to the fast evolution of colors. We note once again that our classification is based on apparent color selections designed to identify the Balmer and D4000 breaks. A passively evolving stellar population develops pronounced Balmer and D4000 breaks in $t\approx 0.3$\,Gyr \citep{Bica94}, while the redshift interval from $z=4.5$ to $z=2.5$ corresponds to $1.2$\,Gyr. This means that a galaxy that quenched its star formation at $z\approx 4-4.5$, and since then has been passively evolving, had enough time to develop the pronounced Balmer and D4000 breaks needed to be identified as quiescent at $z\approx 2.5-3$ through our color selections. Therefore, if the quenching of the star formation occurs at around $z\approx 4-4.5$ for a large number of objects, it is reasonable to expect such a fast evolution in the number of passive galaxies. 
Another possible explanation for the evolution between the two redshift bins is the evolution of the stellar mass: the rapid increase of objects with stellar masses $M_*> 10^{10.6} M_\odot$ may reflect the emergence of a high merging rate, assembling more and more massive quiescent galaxies as they move down with redshift, although the lack of evolution in $M^*$ of the Schechter function can challenge this interpretation.

\subsection{Comparison with models}

According to our results and other results in the literature, the H15 model tends to underpredict quiescent galaxies at $z\gtrsim 2.5$, both in number and in the content of their stellar mass, as visible in Fig.~\ref{dens}. In particular, when considering the bona fide sample, we find that the H15 model underpredicts the number of quiescent objects in the redshift bin $3.0\leq z<3.5$ by a factor of $\sim 4.9\,_{-2.4}^{+2.3}$ up to a value of $\sim 6.0\,_{-3.5}^{+3.1}$ at $z\sim 4$. By checking the differences in the mass densities, we find that the model underpredicts observed values at $3.0\leq z<3.5$ by a factor of $\sim 4.2\,_{-2.4}^{+2.0}$ and it shows a difference by a factor of $\sim 6.2\,_{-2.7}^{+2.5}$ at $z\sim 4$. The differences between H15 model and our results are all the more evident when considering the extended sample. In this case, in the redshift bin $2.5\leq z<3.0$ we find a difference by a factor of $\sim 11.9\,_{-2.9}^{+2.7}$ in the number densities, while the mass densities differ by a factor of $\sim 11.1\,_{-1.2}^{+1.0}$. Moreover, at $z\sim 4$ we find a difference by a factor of $\sim 10.1\,_{-6.1}^{+5.8}$ in the number densities and by a factor of $\sim 19.2\,_{-4.1}^{+3.9}$ when considering the mass densities (also in this case, we simply propagated the errors in our measurements).

Considering the SMFs in the left panel of Fig.~\ref{MF} (i.e., at $2.5\leq z<3.0$), the H15 model appears not to reproduce the shape of the observed SMFs. In particular, it emerges that the model overpredicts low-mass ($\log(M_*/M_\odot)\lesssim 10.4$) and high-mass ($\log(M_*/M_\odot)\gtrsim 11.4$) systems while objects in the intermediate mass range appear to be underpredicted (by a factor of $\sim 8.7$ at $\log(M_*/M_\odot)=10.9$). Also, in the right panel of Fig.~\ref{massfct} (i.e., at $3.0\leq z<4.5$), it is evident that the model seems to under-predict objects in the whole mass range (by a factor of $\sim 4.4$ at $\log(M_*/M_\odot)=10.9$) with the exception of low-mass systems, i.e., $\log(M_*/M_\odot)\lesssim 10.2$.
These discrepancies reflect the intrinsic difficulty in treating the processes involved in galaxy formation, and in particular, the processes related to the transformation of star-forming galaxies into quiescent objects and their mass assembly at $z\gtrsim 2.5$. In  \citet{Cecchi19} we compare high-redshift observed quiescent galaxies to different semi-analytic models (SAMs) and discuss the quenching mechanisms that led to their formation.

In the current scenario, such massive quiescent galaxies are the result 
of a strong active galactic nuclei (AGN) feedback from the central supermassive black holes (BHs) which  significantly affects galaxy formation processes. The AGN feedback can take place in two main ways: the jet (or radio) mode, and the radiative mode (sometimes called quasar mode or bright mode).
The mechanism that is primarily, or entirely, responsible for quenching in the jet mode is connected with highly collimated jets of relativistic particles, in which star formation dies out because the hot gas halo is continually heated, and the supply of new cold gas is cut off \citep{Bower06,Croton06,Somerville08,Kimm09}.
In the radiative mode, quenching is associated with mergers and rapid BH growth, followed by a quasar wind which expels gas from galaxy's center, explaining the growth of the quiescent population \citep{Hopkins08A,Hopkins08B}.  
Cosmological zoom-in simulations, which include fast momentum-driven AGN winds, also appear to be able to quench star formation. Moreover, it appears that it can also maintain quiescence over long timescales without any explicit jet mode type feedback \citep{Choi14}.
Conversely, \citet{Gabor15} suggested that the presence of a gaseous halo kept hot by AGN feedback is sufficient to quench a galaxy without the need for additional radiative mode feedback. 
What the relative importance is of these AGN feedback mechanisms that are responsible for the early appearance of the population of quiescent galaxies at high redshift is still a matter of debate.
However, a feedback mechanism is fundamental in order to quench star formation and form massive quiescent galaxies.

In the H15 model, massive galaxies can be quenched by AGN feedback depending on black-hole and hot-gas mass and, therefore, indirectly on stellar mass. In addition, galaxies of any mass can be quenched by ram-pressure or tidal stripping of gas and through the suppression of gaseous infall. It is argued by \citet{Henriques17} that this combination of processes produces quenching efficiencies which depend on stellar mass, host halo mass, environment density, distance to group centre, and group central galaxy properties. In the case of massive galaxies, the quenching is likely due to AGN feedback. As explained in \citet{Henriques17}, both the quasar and the radio accretion modes on the black hole are considered in the model; while the quasar mode produces no feedback on the galaxy, the radio mode produces a strong feedback by avoiding further hot gas condensation and, therefore, star formation is suppressed once the cold gas is exhausted.

If our result (that is, the under-prediction of quiescent objects at $2.5 \lesssim z\lesssim 3$ in H15) is confirmed, it would imply that the considered model is not efficient enough in producing massive quiescent galaxies at high redshifts. Considering the ingredients of the model, this may be due to the parametrization of the radio mode AGN feedback on the galaxy, or the timescale of this feedback process is too long to produce massive quiescent objects at such redshifts or, in addition, a different accretion mechanism on the black hole should be considered.

On the observations side, the scatter between different observed samples of quiescent objects makes it evident that there is a need for deeper observations on different fields (all the data reported in the plot for quiescent galaxies belong to the COSMOS field) and spectroscopic confirmation to better constrain the population of massive quiescent galaxies at $z\gtrsim 2.5-3.0$, whose presence should be explained by models of galaxy formation but still aren't (see \citealt{Somerville15,Naab17} for two extensive reviews on theoretical state-of-the-art models of galaxy formation). 


\section{Summary and Conclusions}
\label{concl}

The results of the present study can be summarized as follows:
\begin{enumerate}
\item We identified two new color selections using near-infrared bands to select quiescent galaxies at $2.5 \lesssim  z \lesssim  4.5$. The color selections are based on the identification of strong spectral features characterizing these evolved objects, that is, the D$4000$ and Balmer breaks.

\item We studied the effectiveness of our color selections by exploring all the parameters characterizing evolutionary tracks, as well as the effects of emission lines.

\item We applied the color selection to the COSMOS2015 catalog, selecting a parent sample with a cut at $m_{[4.5]}\leq 24$. Through a SED fitting analysis we tightened the selection considering only the most massive ($\log (M_*/M_\odot) >10.6$) high redshift ($z_{\rm{ phot}} \gtrsim 2.5$) quiescent ($\log ({\rm sSFR\, [\,yr^{-1}]}) <-10.5$) galaxies. The proposed color selections coupled with SED fitting analysis allowed us to build a reliable sample of quiescent candidates, aimed at maximizing its completeness.

\item The objects consistent with being the most quiescent massive galaxies (i.e., the bona fide sample) have observed number densities that decrease by a factor of $\sim 1.7$ from the redshift bin $3.0\leq z<3.5$ to the bin $z\geq 4$ while mass densities show almost no evolution with redshift. Considering the extended sample, we found a decrease by a factor of $\sim 12.4^{+3}_{-7}$ in the number of quiescent object and by a factor of $\sim 10.9^{+3}_{-6}$ in their mass densities from the lowest redshift bin (i.e., $2.5\leq z<3.0$) to the highest one (i.e., $z\ge 4.0$). In addition, we find a difference in the number density (mass density) by a factor of $\sim 66\,^{+28}_{-27}$ ($\sim 100\,^{+29}_{-35}$) between our lowest redshift bin (i.e., $2.5\leq z<3.0$) and \citet{Davidzon17} data for the quiescent population in the redshift bin $0.2<z<0.5$.

 \item We estimated the stellar mass functions for our sample of quiescent galaxies and their fit with a Schechter function.

\item According to our results, the semi-analytical model by \cite{Henriques15}  is not able to fully account for the number and mass densities and the stellar mass functions of candidate quiescent objects unless a severe effect of contamination is what is affecting our bona fide sample, in which case the significance of the disagreement with observed data can decrease. In particular, considering the bona fide sample, the model seems to underpredict the number of quiescent objects in the redshift bin $3.0\leq z<3.5$ by a factor of $\sim 4.9\,_{-2.3}^{+2.4}$ that grows to a value of $\sim 6.0\,_{-3.5}^{+3.1}$ at $z\ge 4.0$. In considering the mass densities, we find that the model underpredicts our estimates in the redshift bin $3.0\leq z<3.5$ by a factor of $\sim 4.2\,_{-2.0}^{+2.4}$ and it shows a difference by a factor of $\sim 6.2\,_{-2.5}^{+2.7}$ at $z\ge 4.0$. The differences between H15 model and our results is notably evident when considering the extended sample. In this case, the redshift bin $2.5\leq z<3.0$ demonstrates a difference by a factor of $\sim 11.9\,_{-2.9}^{+2.7}$ in the number densities, while the mass densities differ by a factor of $\sim 11.1\,_{-1.2}^{+1.0}$. Moreover, at $z\ge 4.0$ we find a difference by a factor of $\sim 10.1\,_{-6.1}^{+5.8}$ in the number densities and by a factor of $\sim 19.2\,_{-4.1}^{+3.9}$ considering the mass densities. Moreover, the shape of the SMF is not fully reproduced by the model.

\item This method can be used to select targets for spectroscopic follow-up, especially with future facilities such as the James Webb Space Telescope. If future spectroscopic observations  confirm the presence of a population of quiescent galaxies at high redshift, as proposed in the present work, some further internal mechanisms of quenching will be needed to explain the presence of such galaxies at high redshift that are becoming more and more numerous, given the many studies devoted to this topic. In semi-analytic models, the AGN-feedback needed to quench star-formation at high stellar masses could be an ingredient that still needs to be fully understood and  parametrized differently. 

\end{enumerate}

\begin{acknowledgements}
We thank Bruno Henriques, Gianni Zamorani and Lucia Pozzetti for fruitful discussions on the topic. Data compilations the studies used in this paper were made much more accurate and efficient by the online WEBPLOTDIGITALIZER code. We acknowledge the support from the grants PRIN-MIUR 2015 and ASI n.2018-23-HH.0.
\end{acknowledgements}

\bibliographystyle{aa} 
\bibliography{massive_highz}

\begin{appendix}
\section{Exploring evolutionary tracks's parameters space}\label{appendix1}

We explored the effects of different evolutionary tracks on the choice of color criteria. We analyzed, in particular, the impact of considering star forming galaxies with emission lines, different extinction laws, IMFs, stellar population synthesis (SPS) models, metallicities, SFHs, and redshifts of formation. In all following plots: gray shaded areas represent the selection region for quiescent galaxies, colored numbers represent the redshift of the nearby point for the evolutionary track with the same color and a vector corresponding to a magnitude extinction of $A_V=1$ using Calzetti's law is shown.

\subsection{Emission lines}
\begin{figure*}
\centering
\includegraphics[width=16cm]{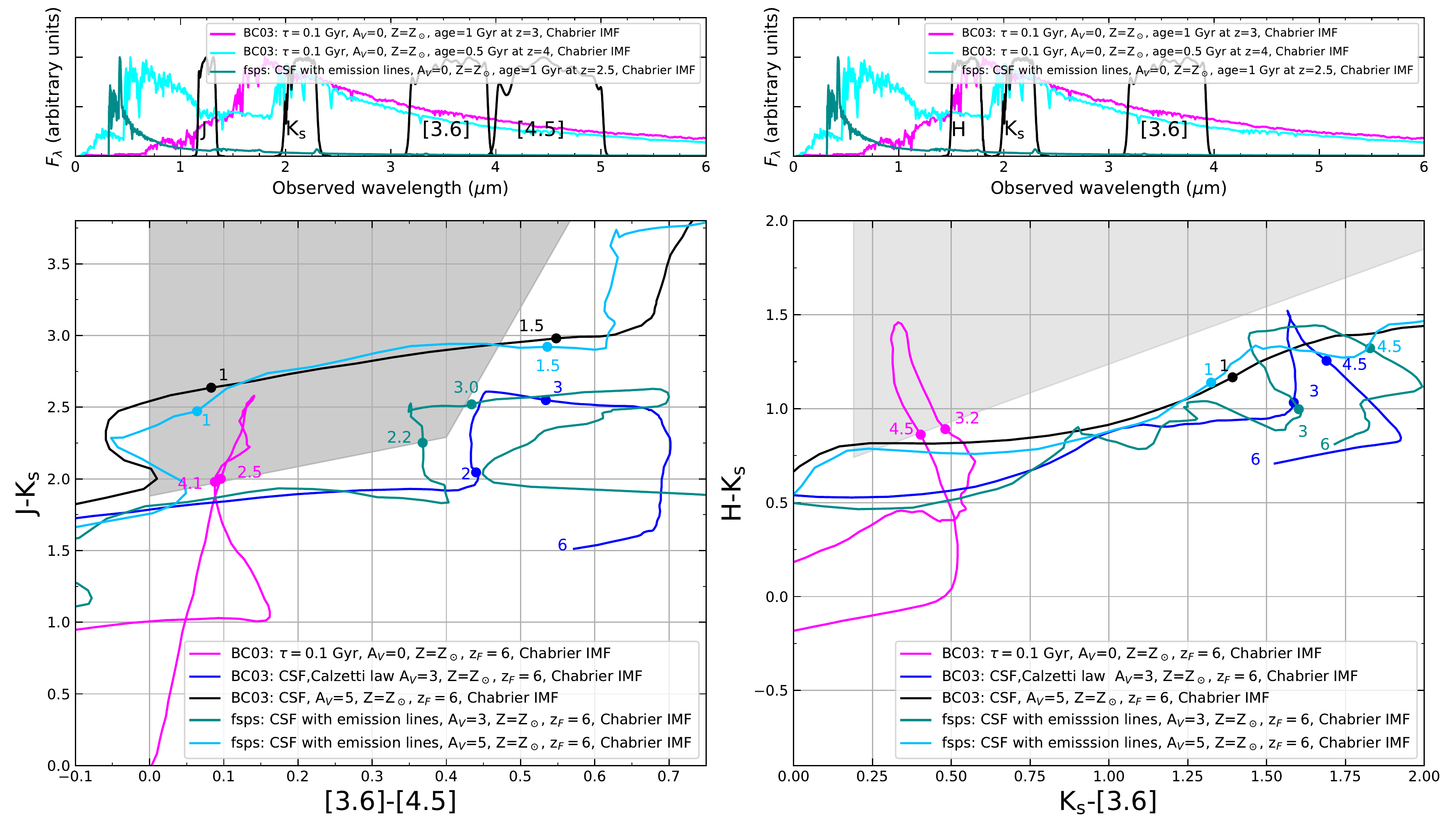}
\caption{Effect of emission lines on $JK_{\rm s}[3.6][4.5]$ and $HK_{\rm s} [3.6]$ color-color diagrams. \textbf{Top panels:} two SEDs of quiescent galaxies (magenta and cyan lines), representing a population 1\,Gyr old redshifted to $z=3$ and a population $0.5$\,Gyr old redshifted to $z=4$ (i.e., $z_{\rm form} \approx 6$) built with BC03 models. A star forming SED with emission lines built from \emph{fsps}, 1 Gyr old and redshifted to $z=2.5$ is shown in dark-cyan. Also the filters transmission curves, used in COSMOS field and in the tracks computation, are shown.  \textbf{Bottom:} evolution in  color-color plots for star-forming and passive galaxies. Tracks representing star-forming galaxies are shown in blue, black dark-cyan, and cyan, and their characteristics are shown in the legend inside the plot. A track for quiescent galaxies (with $e$-folding time of $\tau=0.1$ Gyr and solar metallicity) is shown in magenta.}
\label{fsps}%
\end{figure*}

We built the evolution of a galaxy template of constant star formation with the contribution of nebular emission lines using the code \emph{fsps} \citep{Conroy09,Conroy10}. As shown in Fig.~\ref{fsps}, considering a template with emission lines in the $HK_{\rm s} [3.6]$ diagram does not affect the selection criteria. Instead, considering the $JK_{\rm s}[3.6][4.5]$ selection, some contamination by star forming galaxies with emission lines is expected at $2.2\lesssim z\lesssim 3.0$ for extinction values around $A_V\sim 3$. In Fig.~\ref{fsps}, it can be seen that higher (or lower) values of $A_V$ would not affect the selection box contaminating the candidates. Considering the SED fitting procedure for a star-forming template with emission lines, this contamination may be kept under control with the aid of optical bands where the emission-line galaxies should be characterized by a larger flux with respect to a quiescent object (see also the top panels of Fig.~\ref{fsps}).

\subsection{Extinction laws}
\label{sec_extlaws}

\begin{figure*}
\centering
\includegraphics[width=16cm]{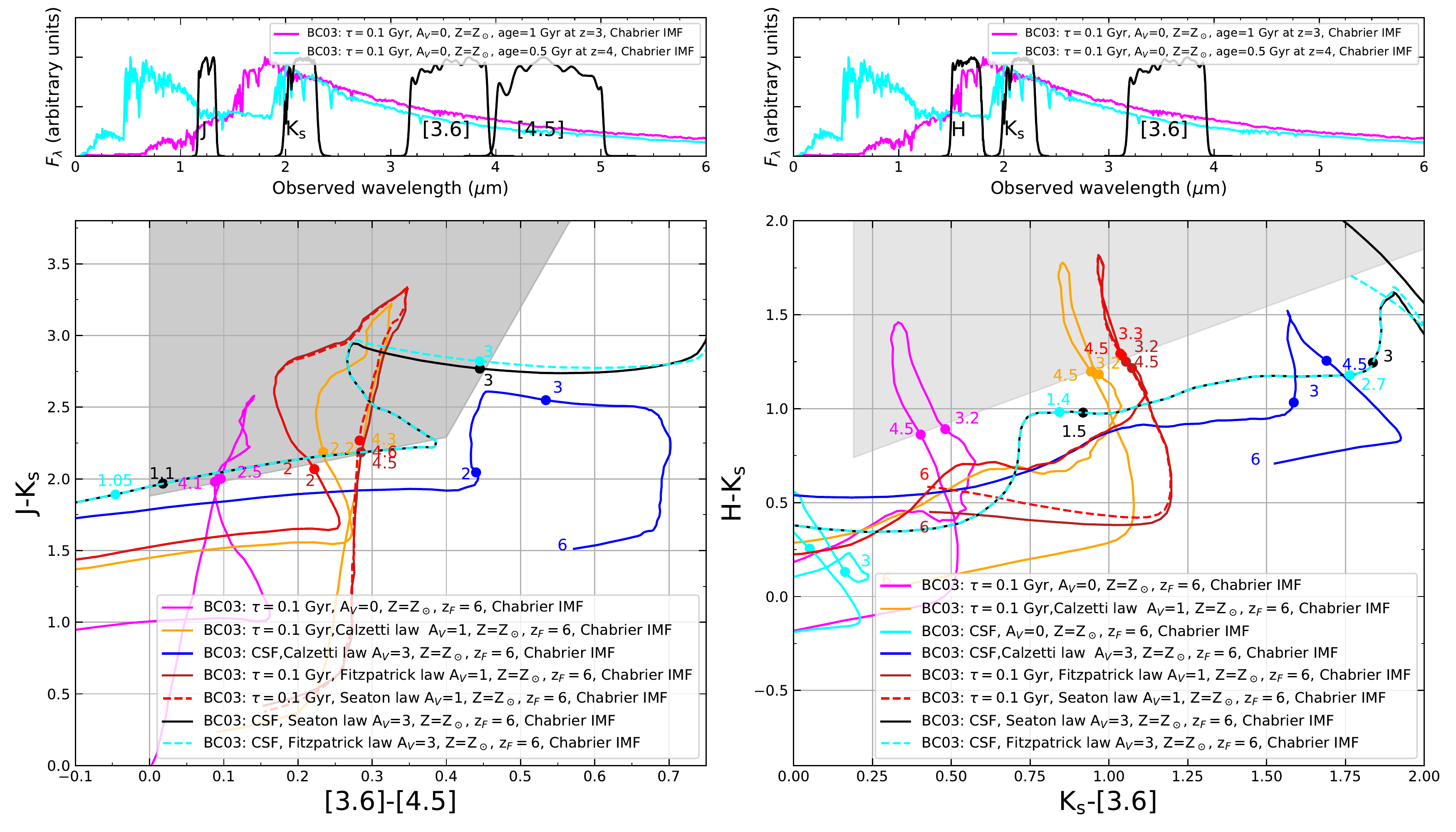}
\caption{Effect of different extinction laws on $JK_{\rm s}[3.6][4.5]$ and $HK_{\rm s} [3.6]$ color-color diagrams. \textbf{Top panels:} two SEDs of quiescent galaxies (magenta and cyan lines), that is, a population 1 Gyr old redshifted to $z=3$  and a population $0.5$ Gyr old redshifted to $z=4$ (i.e., $z_{\rm form} \approx 6$) built with BC03 models.  \textbf{Bottom:} evolution in the color-color plots for star-forming and passive galaxies. Tracks representing star-forming galaxies are shown in solid blue, solid black and dashed cyan lines, for the Calzetti, Seaton and Fitzpatrick laws, respectively, and their characteristics are shown in the inserted legends. Tracks for quiescent galaxies (with $e$-folding time of $\tau=0.1$ Gyr and solar metallicity) are shown in solid magenta, solid orange, dashed red  and brown for the Calzetti ($A_V=0$ and $A_V=1$), Seaton and Fitzpatrick laws, respectively.}
\label{extlaws}%
\end{figure*}

We also tested our color-color diagrams using three different extinction laws. In particular, as shown in Fig.~\ref{extlaws}, we computed the evolutionary tracks using the Calzetti law \citep{Calzetti00}, characteristic of starburst (SB) galaxies, the Fitzpatrick law \citep{Fitzpatrick86} derived for the Large Magellanic Cloud (LMC), and the Seaton law \citep{Seaton79} obtained for the Milky Way (MW). The main differences among them are the bump at $2200$\,\AA, which is absent in starburst galaxies, and the slope in the UV, which is steeper for the LMC extinction law. 

As shown in Fig.~\ref{extlaws}, tracks built with MW and LMC laws are, in general, very similar and differ slightly from the SB law we used as a reference in the present paper. Adopting different extinction laws does not affect the selection criteria of quiescent galaxies.  The tracks of color evolution for star-forming galaxies can cross our selection boxes in the $JK_{\rm s}[3.6][4.5]$ diagram when adopting LMC and MW extinction laws, but only at $z \lesssim 3$.

\subsection{IMF}

We tested our selections with the IMFs of \citet{Salpeter55,Chabrier03,Kroupa01} and we found no appreciable differences in evolutionary tracks . This was expected because the SED of a passively evolving galaxy is mainly determined by the stars at the turnoff and considering the possible ages of a galaxy at $z\ga 3,$ the stars dominating the emission, and, therefore, the colors, are in the portion of the IMF that is similar for all the mentioned IMFs. 

\subsection{SPS models}

\begin{figure*}
\centering
\includegraphics[width=16cm]{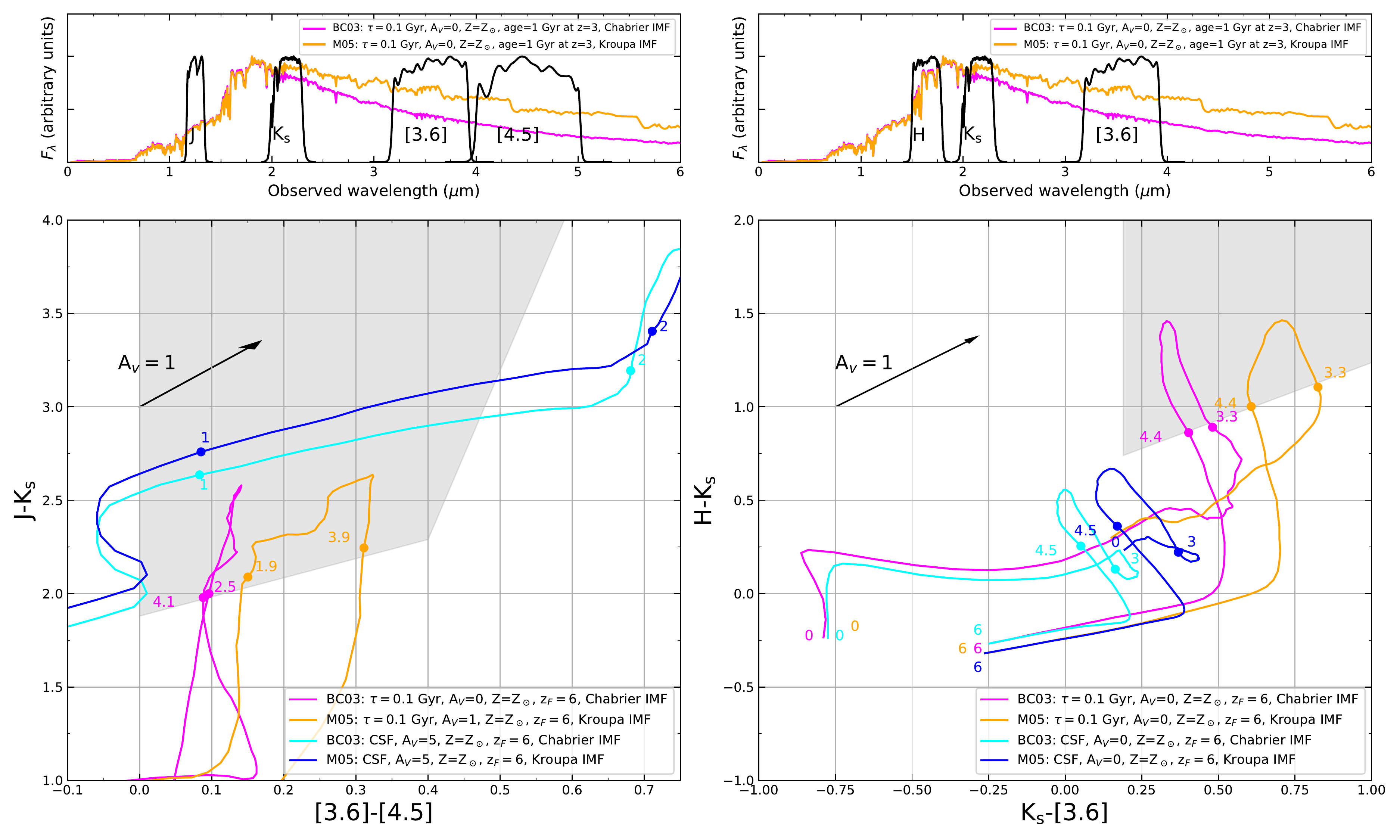}
\caption{Effect of the choice of SPS models on $JK_{\rm s}[3.6][4.5]$ and $HK_{\rm s} [3.6]$ color-color diagrams. \textbf{Top panels:} two SEDs of quiescent galaxies, that is, a population 1 Gyr old redshifted to $z=3$ (i.e. $z_{\rm form} \approx 6$) built with BC03 models (magenta line) and with M05 models (orange line). \textbf{Bottom:} evolution in color-color plots for star-forming and passive galaxies. Tracks representing star-forming galaxies are shown in blue and cyan, while their characteristics are shown in the inserted legends. Tracks for quiescent galaxies (with $e$-folding time of $\tau=0.1$ Gyr and solar metallicity) are shown in magenta and orange for BC03 and M05, respectively.}
\label{models}%
\end{figure*}

The difference between BC03 and M05 models \citep{B&C03,Maraston05} for the building of evolutionary tracks is explored in Fig.~\ref{models}. 
As explained in Sect~\ref{BC03vsM05}, for $\lambda >2-2.5\,\mu$m and ages between  $\sim 0.2$ and $\sim 2$ Gyr, a notable difference in flux between the two models is expected \citep{Maraston06} which means that a difference in $K_{\rm s}-[3.6]$ color is also expected. 
However, both models generate evolutionary tracks for quiescent galaxies which fall in the selection region for approximately the same redshift range, and leave tracks for star-forming galaxies outside of it. Therefore, the choice of the model adopted does not affect the color selection criteria.
The difference between BC03 and M05 results in SED fitting for observed galaxies is also explored in Sect.~\ref{BC03vsM05}, where no appreciable difference has been found.

\subsection{Metallicity}

\begin{figure*}
\centering
\includegraphics[width=16cm]{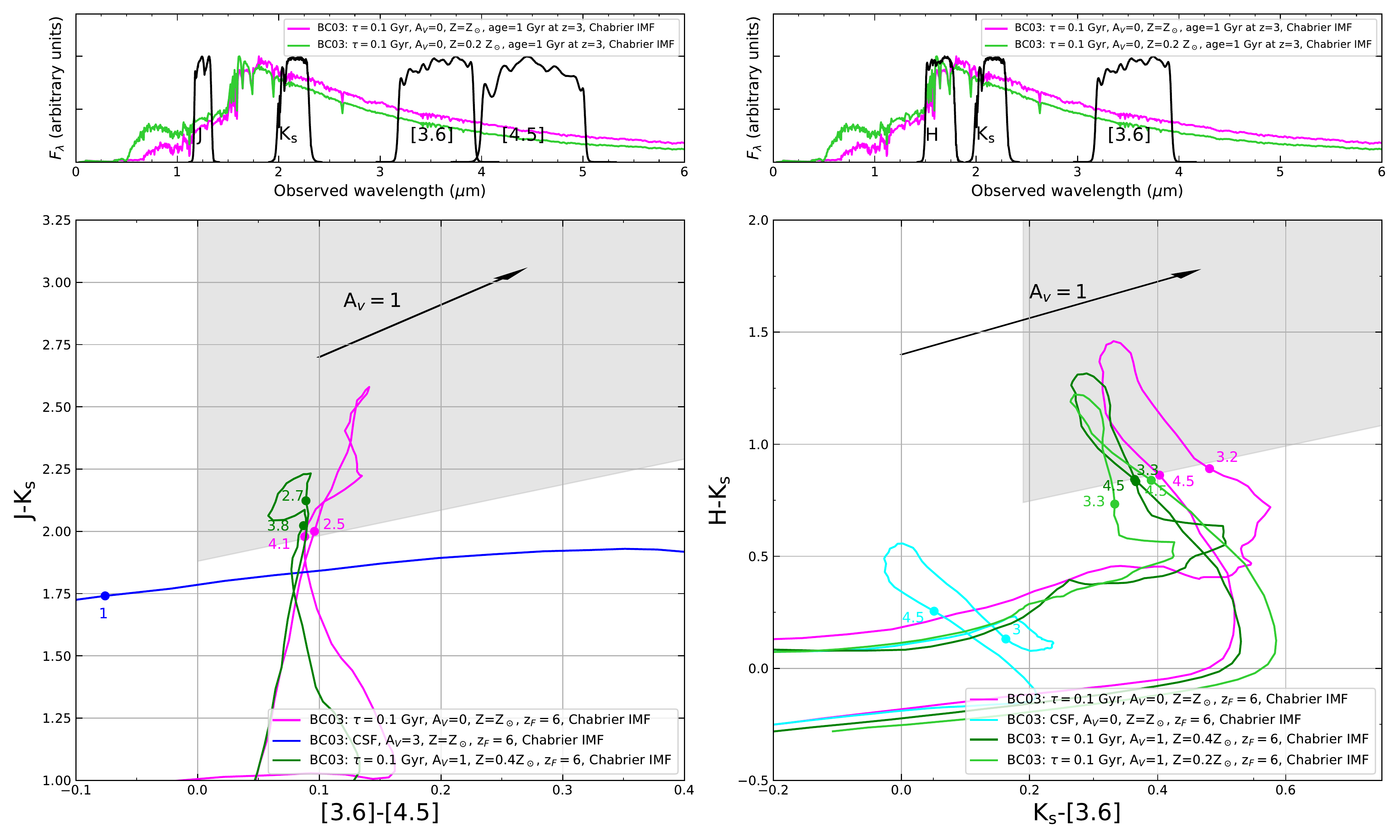}
\caption{Effect of metallicity on $JK_{\rm s}[3.6][4.5]$ and $HK_{\rm s} [3.6]$ color-color diagrams. \textbf{Top panels:} two SEDs of quiescent galaxies are shown. In particular, the magenta SED represents a population 1 Gyr old redshifted to $z=3$ (i.e. $z_{\rm form} \approx 6$) with $Z=Z_\odot$, while the green line shows the SED of a population of a 1 Gyr redshifted to $z=3$ (i.e. again $z_{\rm form}\approx 6$) with $Z=0.2\, Z_\odot$. \textbf{Bottom:} color-color plots with evolutionary tracks for star-forming and passive galaxies. Tracks representing star-forming galaxies are shown in blue and cyan, while their characteristics are shown in the inserted legends. Three tracks for quiescent galaxies are shown in magenta, dark-green and light-green for different metallicities ($Z=Z_\odot$, $Z=0.4Z_\odot$, $Z=0.2Z_\odot$ respectively).}
\label{Z}%
\end{figure*}

The effect of different metallicities is explored in Fig.~\ref{Z} (for $Z=0.02\, ,\, 0.008\, ,\, 0.0004$): we found that all the derived tracks enter the color selection for $z\approx 3-3.5$ and exit for $z\approx 4-4.5$, suggesting that the color selection is valid for a wide range of metallicities . Assuming a metallicity in the range of $Z=Z_\odot$ to $0.2\,\mathrm{Z}_{\odot}$ is reasonable  even for high redshift galaxies (e.g., \citealt{Sawicki1998,Papovich2001,Shapley2001,Mannucci10,Finkelstein16}). For instance, according to \citet{Glazebrook17}, a quiescent galaxy at $z=3.7$ is consistent with having a solar metallicity. 

\subsection{Star formation histories}

As already mentioned, star formation histories for quiescent galaxies have been parametrized with an exponentially declining star formation with short $e$-folding times in order to guarantee a negligible SFR after a few hundreds Myr. Star formation histories follow the formula ${\rm SFR}(t)\propto \tau^{-1} e^{-t/\tau}$ where $\tau$ is the $e$-folding time. High-redshift quiescent galaxies need to have formed their stellar mass quickly since they are observed at $z\ge3$ when the universe was only $2$ Gyr old; therefore, a timescale of $\tau=0.1$ Gyr is often assumed. By considering different $e$-folding times and assuming the same redshift of formation $z_{\rm form}=6$, models with a longer $e$-folding time will be characterised by bluer colors with respect to the model, with a shorter $\tau$ at the same redshift. The color selection is still valid assuming an $e$-folding time of $0.3$ Gyr, even though at fixed ages (or redshift) tracks with longer $e$-folding times are bluer. Assuming even larger $e$-folding times will not allow for the selection of passive galaxies at high redshift since the stellar population do not have enough time to evolve and develop the D$4000$ and Balmer breaks. 

Delayed star formation histories, parametrized as $\propto \tau^{-2} t \, e^{-t/\tau}$, have also been explored. In particular, choosing timescales of the SFH that are small compared to the time between $z_{\rm form}$ and $z_{\rm obs}$ (e.g., $\tau\approx 0.1$ Gyr), the color selections are still valid, although on a slightly narrower redshift interval. Longer $e$-folding times would instead represent star-forming galaxies in the redshift interval $z\approx 2.5-4$.

\begin{figure*}
\centering
\includegraphics[width=16cm]{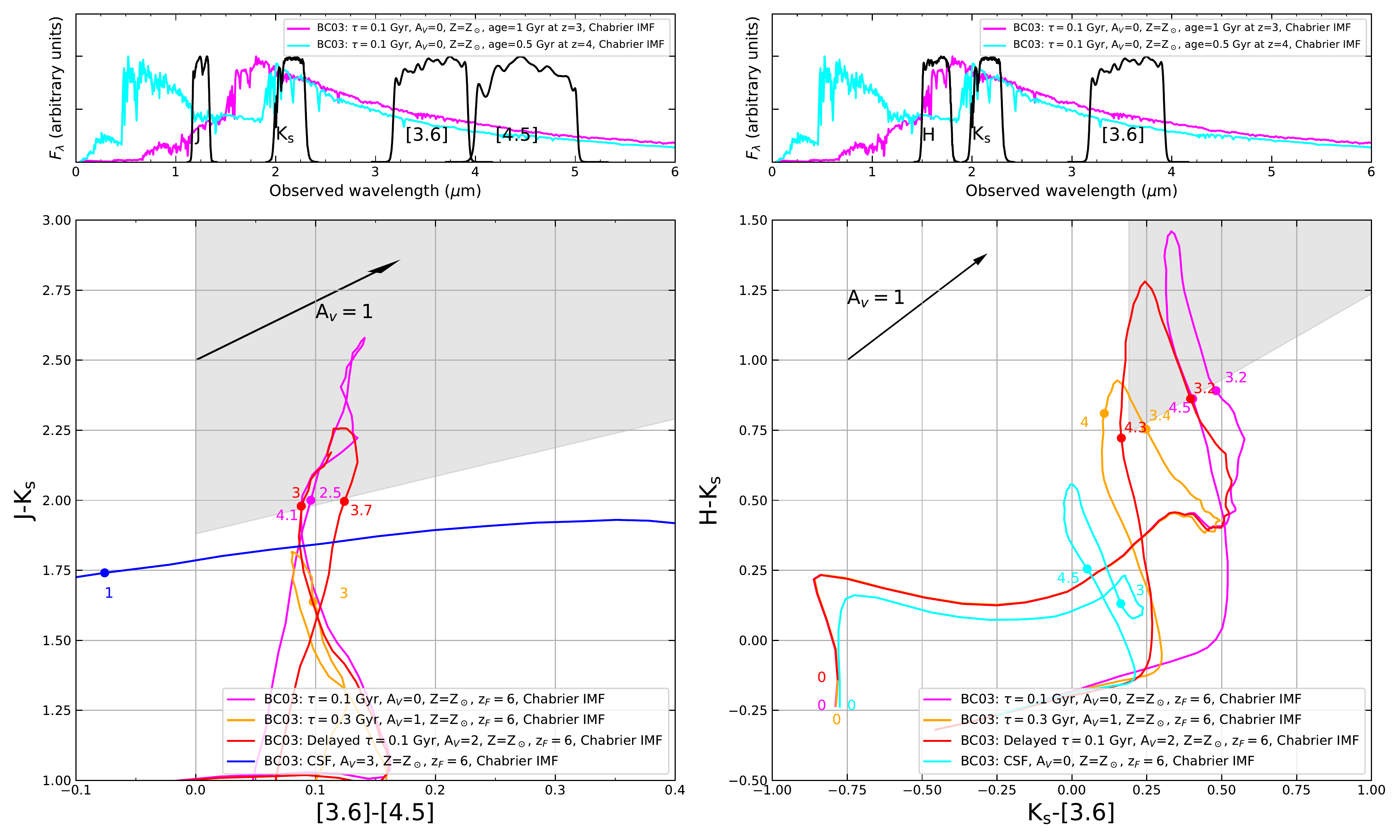}
\caption{Effect of SFHs on $JK_{\rm s}[3.6][4.5]$ and $HK_{\rm s} [3.6]$ color-color diagrams. \textbf{Top panels:} two SEDs of quiescent galaxies are shown in magenta SED and cyan, representing a population of $1$ Gyr redshifted to $z=3$ (i.e. $z_{\rm form} \approx 6$), and a population of 0.5 Gyr redshifted to $z=4$ (i.e., again $z_{\rm form}\approx 6$), respectively. \textbf{Bottom:} color-color plots with evolutionary tracks for star-forming and passive galaxies. Tracks representing star-forming galaxies are shown in blue and cyan and their characteristics are shown in the inserted legends. Three tracks for quiescent galaxies are shown in magenta, orange and red: the magenta track is parametrized with an exponentially declining star formation with $\tau=0.1$ Gyr, orange track has $\tau=0.3$ Gyr, and red track has been parametrized with a delayed star formation history where $\tau=0.1$ Gyr.}
\label{sfh}%
\end{figure*}

\subsection{Redshift of formation}
\label{sect_zf}

Models with higher $z_{\mathrm{form}}$ are, in general, characterized by redder colors at the same redshift since the galaxy which forms at a higher $z$ will exhibit prominent breaks at earlier epochs. This leads to a slightly wider redshift interval for the selection (for $z_{\mathrm{form}}=8$ we select quiescent galaxies in the range $3\lesssim z\lesssim 4.8$ for $HK_{\rm s} [3.6]$ and in the range $2.5\lesssim z\lesssim 4.5$ for $JK_{\rm s}[3.6][4.5]$) since galaxies enter the color selections at higher redshifts than they did with our adopted reference value of $z_{\rm form}=6$.

\begin{figure*}
\centering
\includegraphics[width=16cm]{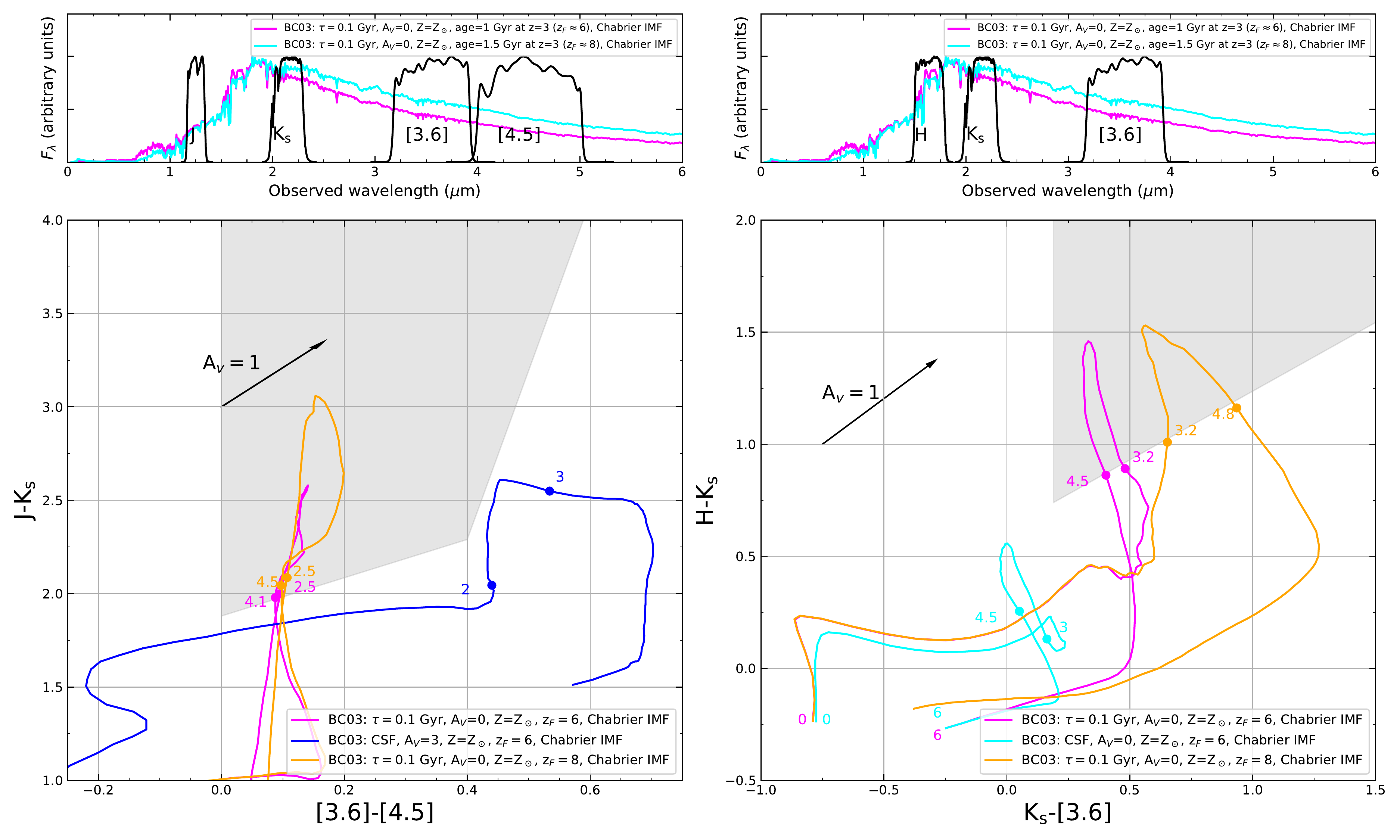}
\caption{Effect of formation redshift on $JK_{\rm s}[3.6][4.5]$ and $HK_{\rm s} [3.6]$ color-color diagrams. \textbf{Top panels:} two SEDs of quiescent galaxies are shown. In particular,  magenta SED represents a population of 1 Gyr redshifted to $z=3$ (i.e. $z_{\rm form} \approx 6$), while  cyan line shows SED for a population of 1.5 Gyr redshifted to $z=3$ (i.e. $z_{\rm form}\approx 8$). \textbf{Bottom:} color-color plots with evolutionary tracks for star-forming and passive galaxies.  Tracks representing star-forming galaxies are shown in blue and cyan and their characteristics are shown in the inserted legends. Two tracks for quiescent galaxies with redshift of formation $z_{\rm form}=6$ and $z_{\rm form}=8$ are shown in magenta and orange respectively. Their $e$-folding time is $\tau=0.1$ Gyr and they have solar metallicity.}
\label{zf}%
\end{figure*}


\end{appendix}

\end{document}